\documentclass[a4,twocolumn]{aa}

\usepackage{graphicx,txfonts,natbib,verbatim}
\bibpunct{(}{)}{;}{a}{}{,}

\begin{document}

\title{Differential Emission Measures from the \\Regularized Inversion of Hinode
and SDO data}

\author{I. G. Hannah \and E. P. Kontar}

\offprints{Hannah \email{iain.hannah@glasgow.ac.uk}}

\institute{SUPA School of Physics \& Astronomy, University of Glasgow,
Glasgow, G12 8QQ, UK}

\date{Received ; Accepted }

\abstract{}{To demonstrate the capabilities of regularized inversion to recover
Differential Emission Measures (DEMs) from multi-wavelength observations
provided by telescopes such as Hinode and SDO.}{We develop and apply
an enhanced regularization algorithm, used in RHESSI X-ray spectral analysis, to
constrain the ill-posed inverse problem that is determining the DEM from solar
observations. We demonstrate this computationally fast technique applied to a
range of DEM models simulating broadband imaging data from SDO/AIA and
high resolution line spectra from Hinode/EIS, as well as actual active region
observations with Hinode/EIS and XRT. As this regularization method naturally
provides both vertical and horizontal (temperature resolution) error bars we are
able to test the role of uncertainties in the data and response functions.}
{The regularization method is able to successfully recover the DEM from
simulated data of a variety of model DEMs (single Gaussian, multiple Gaussians
and CHIANTI DEM models). It is able to do this, at best, to over four orders of
magnitude in DEM space but typically over two orders of magnitude from peak
emission. The
combination of horizontal and vertical error bars and the regularized solution
matrix allows us to easily determine the accuracy and robustness of the
regularized DEM. We find that the typical range for the horizontal errors is
$\Delta$log$T\approx 0.1 -0.5$ and this is dependent on the observed signal to
noise, uncertainty in the response functions as well as the source model and
temperature. With Hinode/EIS an uncertainty of 20\% greatly broadens the
regularized
DEMs for both Gaussian and CHIANTI models although information about the
underlying DEMs is still recoverable. When applied to real active region
observations with Hinode/EIS and XRT the regularization method is able to
recover a DEM similar to that found via a MCMC method but in considerably
less computational time.} {Regularized inversion quickly determines the DEM
from solar observations and provides reliable error estimates (both horizontal
and vertical) which allows the temperature spread of coronal plasma to be
robustly quantified.}

\keywords{Sun:Corona - Sun:Flares - Sun: X-rays, gamma rays - Sun:activity -
Sun:UV radiation}
\titlerunning{Regularized Inversion of Hinode and SDO data}

\authorrunning{Hannah \& Kontar}
\maketitle

\section{Introduction}

Observations of the solar atmosphere with temperature sensitive spectral lines
provide crucial information about the temperature distribution of the emitting
plasma. These are vital for trying to resolve the question of which mechanisms
heat different solar phenomena. Such as whether coronal loops are heated
by a nanoflare model of magnetically reconnecting multi-braided loop strands
\citep[e.g.][]{1988ApJ...330..474P} or chromospheric evaporation
\citep[e.g.][]{1974SoPh...34..323H}. Or is the hot emission observed in large
through to micro- flares \citep{2011SSRv..159...19F,2011SSRv..159..263H}
predominantly due to energetic particles or other mechanisms such as
waves. To reliably answer these questions, one needs to know not only the
uncertainties on the emission for a given temperature, but also the
uncertainties on the temperature itself, i.e. the temperature resolution.

\begin{figure}\centering
\includegraphics[width=60mm]{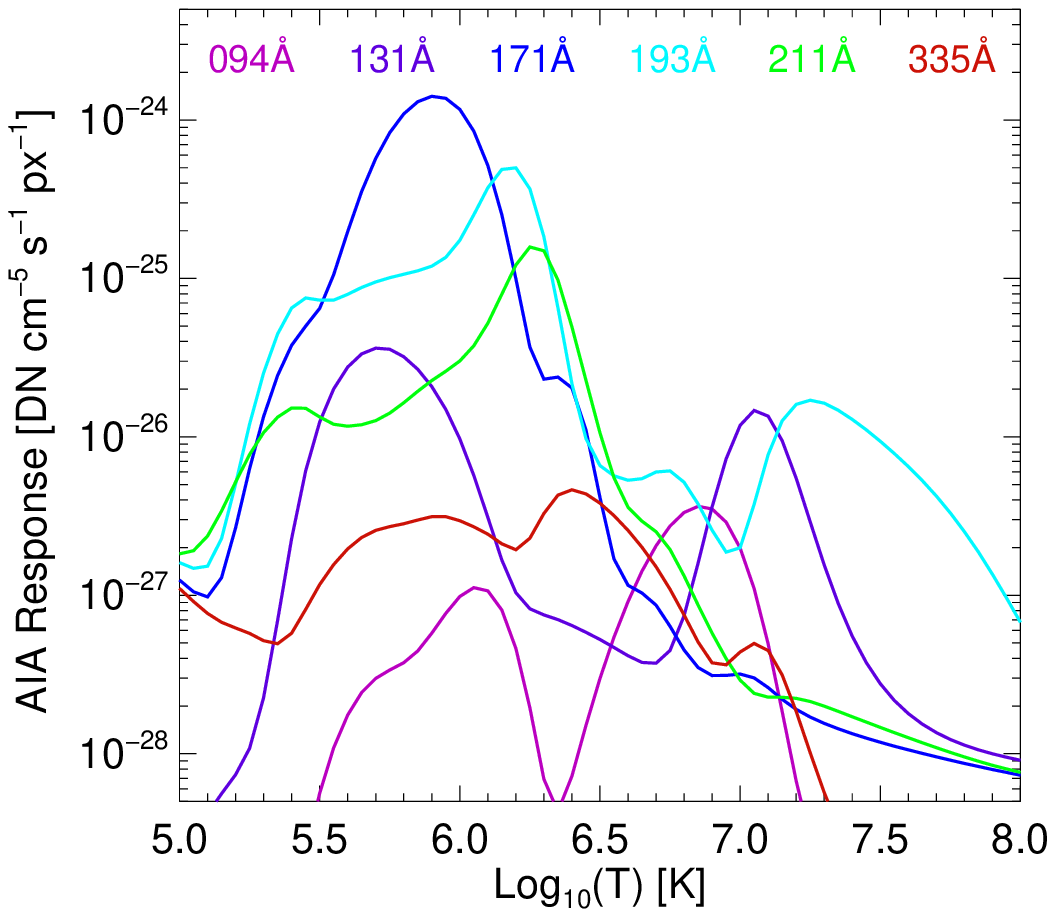}\\
\includegraphics[width=60mm]{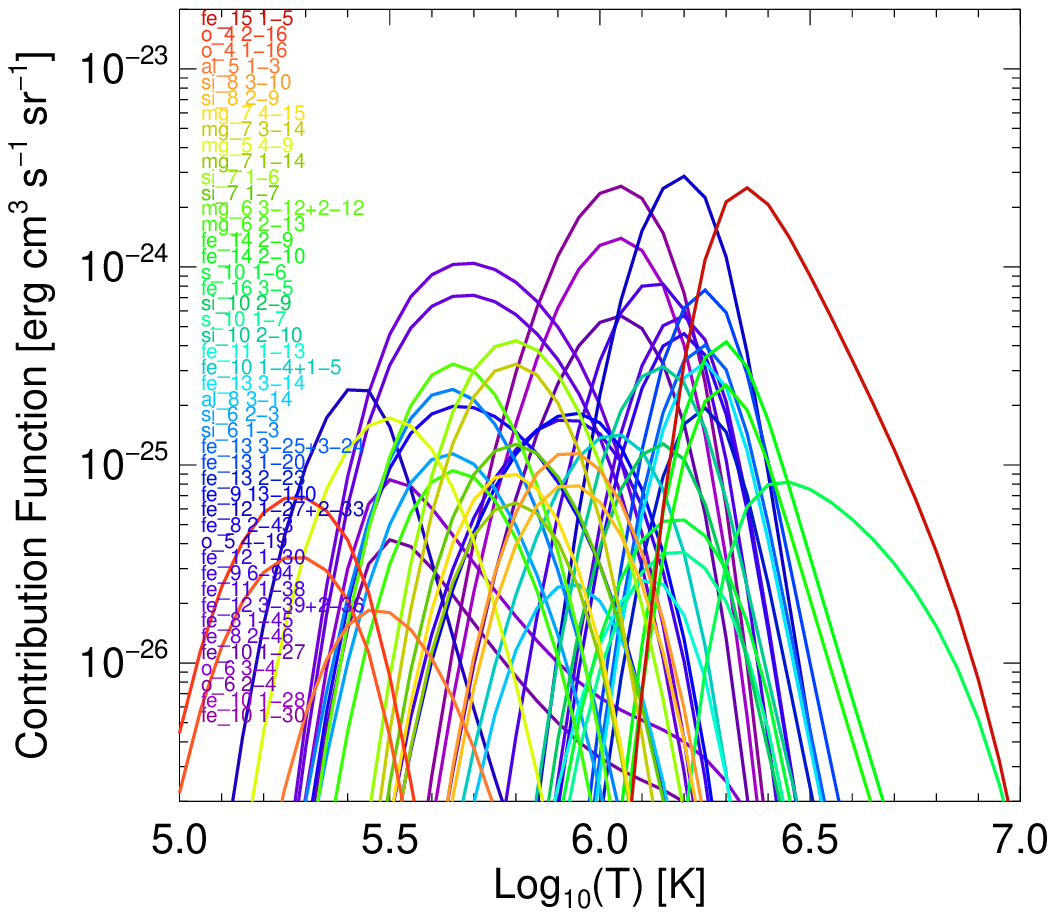}
\caption{\label{fig:tr}Temperature response for SDO/AIA \citep{Boerner}
and the contribution functions for several EUV lines from CHIANTI observable by
Hinode/EIS \citep{2009ApJ...706....1L}.}
\end{figure}

The observations made with Hinode's X-ray Telescope XRT
\citep{2007SoPh..243...63G} and EUV Imaging Spectrometer EIS
\citep{2007SoPh..243...19C} and SDO's Atmospheric Imaging Array AIA
\citep{lemen} and EUV Variability Experiment EVE \citep{woods} provide a
wealth of information about the solar emission over a broad range of
temperatures. Assuming this UV/EUV/X-ray emission is both optically
thin and in thermal equilibrium, via collisions, then the temperature
distribution of plasma emitting along the line of sight $h$ can be described by
the differential emission measure DEM, typically  given by $ \xi(T)=n^2 dh/dT$
[cm$^{-5}$K$^{-1}$] where $n(h(T))$ is the electron density at $h$ and with
temperature $T$ (see Chapter 4 \citet{1992str..book.....M} for detailed
discussion of the different DEM forms). This however, can not be immediately
inferred from such multi-wavelength observations as the DEM is convolved by the
emission processes and the instrumental response, i.e.

\begin{equation}\label{eq:dem_int}
g_i=\int_T K_i(T) \xi(T) dT +\delta g_i
\end{equation}

\noindent where $g_i$ is our observable for the $i^\mathrm{th}$ filter, which
has a temperature dependent response function $K_i(T)$ and $\delta g_i$ is the
error. For spectroscopic
observations this is respectively the line intensity and contribution function
(examples shown in Figure \ref{fig:tr}). The uncertainties associated with these
observations (counting statistics, background and instrumental errors) compounds
the difficulty in the determination of the DEM and results in an ill-posed
inverse
problem
\citep{ti63,1985InvPr...1..301B,1986ipag.book.....C,1996ApJ...457..882S}. Any
direct attempts to solve Eq. \ref{eq:dem_int} normally leads to the
amplification of the uncertainties, and hence spurious solutions.

To reconstruct the DEM additional information (i.e. constraints) has to
be added and numerous approaches have been developed to solve this problem
(for overviews see, for instance, \citet{1991AdSpR..11..281M} or Chapter 5
\citet{2008uxss.book.....P}). The simplest of these is to assume that all
the emission is at a single temperature (isothermal)
with $\xi(T)\propto\delta(T-T_0)$
where $\delta(x)$ is the Dirac delta function. The ratio of emission between two
filters is then equal to the ratio of the response functions at the isothermal
temperature \citep[e.g.][]{2005ApJ...635L.101W,2009ApJ...698..756R}.  Dividing
the observable by the response function $g_i/K_i$ and plotting this as a
function of temperature, the intersect point of the different curves (EM
loci curves) will give the isothermal temperature and emission measure
\citep[e.g.][]{2011ApJ...731...49S}. Although this is a simple and computationally
fast method it does require the isothermal assumption and if the DEM is
multi-thermal this method will produce erroneous results.

Another approach is to forward fit a chosen model, minimising the
differences in observable space. This has been implemented for a discretised
spline model DEM
\citep{1992MmSAI..63..767M,1996ApJS..106..143B,2000A&A...363..800P} and
more recently using the IDL mpfit routine from SDO/AIA and Hinode/XRT in M.
Weber's \texttt{xrt\_dem\_iterative2.pro}\footnote{Available in SolarSoftWare
with the Hinode/XRT software
\$SSW/hinode/xrt/idl/util/xrt\_dem\_iterative2.pro}.
\citep{2004IAUS..223..321W,2004ASPC..325..217G}. This
iterative forward fitting approach has also been developed with multiple
Gaussian model DEMs using the IDL POWELL routine
\citep{2011ApJ...732...81A}. To estimate the error in the DEM with these
methods a Monte Carlo approach is adopted, producing multiple realisations
within a given noise range. These approaches will find parameters for the model
DEM but requires an assumed model and can be computationally slow when
error estimates are required. The Metropolis MCMC approach has also been
used on recovering the DEM for a large set of EUV spectral lines
\citep{1998ApJ...503..450K},  part of PINTofALE spectral analysis package. This
will give a robust measure of the parameter probability space but again can be
computationally intensive, especially if a large number of lines and model
parameters are considered. Bayesian formalism has also been recently
used in a Bayesian Iterative Method (BIM), successfully reconstructing DEMs
from both simulated and observed data \citep{2010A&A...523A..44G}. The
iterative mpfit method was compared to a maximum likelihood and a genetic
algorithm technique, finding similar results between the approaches
\citep{2008AnGeo..26.2999S}. Another genetic algorithm approach, which
involves a preconditioning step where the optimum subset of spectral lines are
selected, was found to be more effective \citep{2000ApJ...529.1115M}.
Currently a Singular Value Decomposition (SVD) inversion approach is also under
development by Weber for SDO/AIA data.

Regularized inversion methods introduce an additional ``smoothness'' to
constrain the amplification of the uncertainties, allowing a stable inversion to
recover the DEM solution
\citep[e.g.][]{1977A&A....61..575C,1986ipag.book.....C}. This was demonstrated
to have promise for solar observations by \citet{1977A&A....61..575C} and
subsequently tested on simulated data by \citet{1991AdSpR..11..281M}. Several
forms of regularized inversion -- truncated (or ``zeroth-order'') SVD,
second-order regularization and maximum entropy regularization -- have also
been tested using simulated EUV spectral line emission
\citep{1997ApJ...475..275J}. Although they determined that these approaches
were superior to the simple ratio method they found several problems with the
regularized inversion:  the smoothness criterion used may not be physically
``appropriate''; the solutions are highly sensitive to uncertainties in the kernel
(${\bf K}$, response or contribution functions); the return of negative solutions.

In this paper we present a regularization method, which resolves some
of these problems and robustly recovers the underlying DEM with
errors\footnote{The code written in IDL requiring SSW is available online:
http://www.astro.gla.ac.uk/$\sim$iain/demreg/}. The method not only
determines the DEM quickly, and its associated errors, but also naturally
provides an estimate to the temperature resolution of the method. The ability to
quickly compute (via Generalised Singular Value Decomposition) the DEM and
its associated errors is due to the fact this method is linear, unlike those using
maximum entropy for example. This method has already been implemented and
applied to solar data for the inversion of RHESSI \citep{2002SoPh..210....3L}
X-ray spectra to their source electron distribution
\citep{2004SoPh..225..293K}. Several other inversion techniques have been
developed to infer X-ray photon
spectra and/or electron spectra from RHESSI data
\citep{2003ApJ...595L.127P,2004ApJ...613.1233M,2006ApJ...643..523B} however
it is the regularized inversion that has become the de facto approach.
It has subsequently been used to infer the DEM, as well as the
non-thermal emission, from RHESSI hard X-ray spectra
\citep{2006SoPh..237...61P}.

\begin{figure*}\centering
\includegraphics[width=55mm]{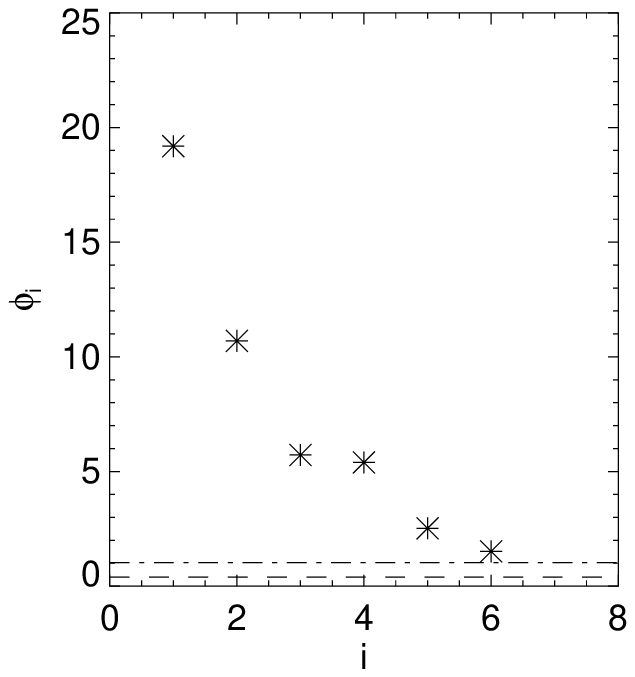}
\includegraphics[width=55mm]{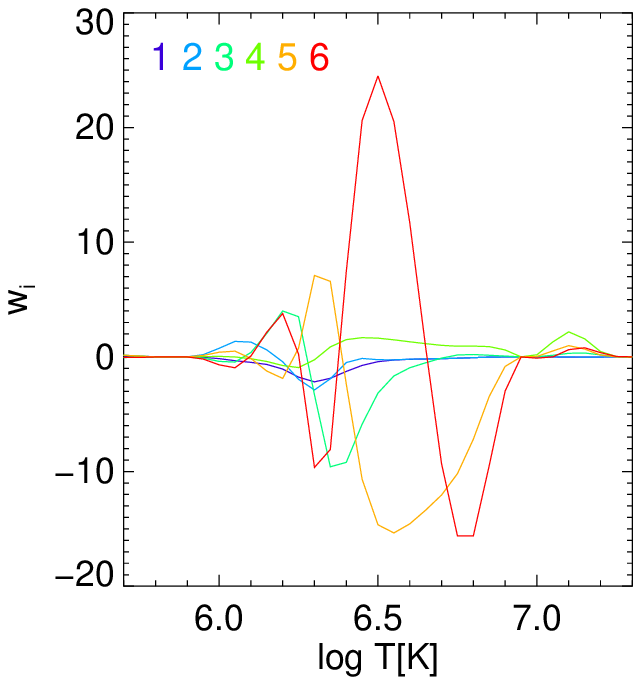}
\includegraphics[width=55mm]{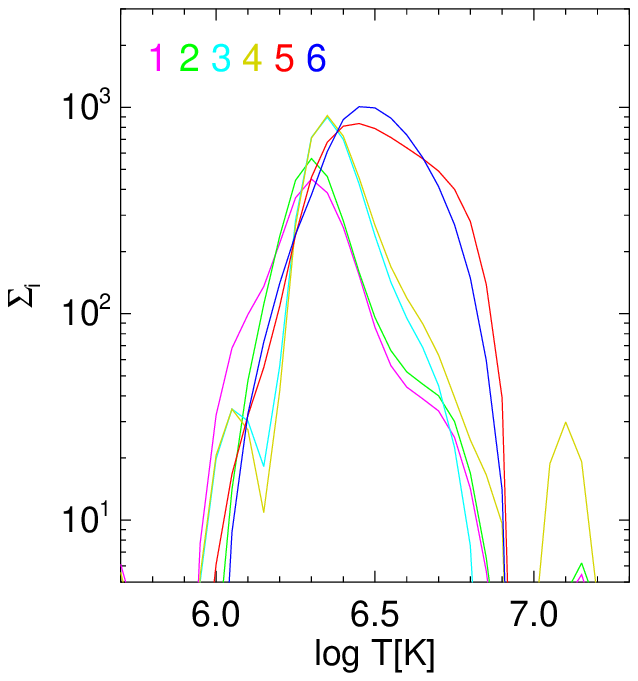}\\
\caption{\label{fig:solution_components}
 Singular values  $\phi_i$ (left), singular vectors ${\bf w}_i$ (middle)
and
 the regularized solution Eq. \ref{eq:sol_gen} as a function of $i$ (right).
The
lines in the singular value plot (left) indicate the regularization parameter
$\lambda$ that produces the solution with desired $\chi^2$ (dashed line) and
with the additional positivity constraint (dash dot line). These values are
from the regularized solution shown in Figure \ref{fig:aia_exae}.}
\end{figure*}

In \S2 we detail the regularization method and how the errors and temperature
resolution are determined. In \S3 and \S4 we demonstrate the
capabilities of the regularization method, in comparison to other methods, on
simulated SDO/AIA and Hinode/EIS data for a variety of model DEMs (single and
multiple Gaussians and the CHIANTI DEM models). In \S5 we use the
regularization method to recover the DEMs from observations of an active
region made with Hinode/EIS and XRT \citep{2010ApJ...711..228W}.

\section{Regularized inversion of multi-wavelength data}\label{sec:reg}
To find the line of sight DEM $\xi(T_j)$   $j=1,..,M$ [cm$^{-5}$ K${^{-1}}$] is
to solve the system of linear equations:

\begin{equation}\label{eq1}
g_i = {\bf K}_{i,j}\;\xi(T_j)
\end{equation}

\noindent where $g_i$ is the observable (either the imaged DN or the
integrated line
intensity) for the specific filter or wavelength $i$ ($i=1,...,N$) and ${\bf
K}_{i,j}$
is the corresponding temperature response or spectral line contribution function
(examples shown in Figure \ref{fig:tr}). Eq. \ref{eq1} is a generally
ill-posed inverse problem and hence the least square problem

\begin{equation}\label{mproblem}
\left\|\frac{{\bf{K}}\;{\xi(T)-{\bf g}}}{\delta {\bf g}}\right\|^2=\mbox{min},
\end{equation}

\noindent does not have a unique solution. In the case $M=N$, a formal
solution of Eq. \ref{eq1} can always be written as $\xi(T_j)={\bf
K}_{i,j}^{-1}g_i
$, where ${\bf K}^{-1}$ is the inverse of matrix ${\bf K}$. However, due to the
large condition number, which is the ratio of the largest to the smallest singular
values of ${\bf K}$, such a solution is meaningless in all practical cases due to
substantial noise amplification \citep{1977A&A....61..575C,1985InvPr...1..301B}.
In addition, with $M>N$ the system is also under-determined. With inherent
statistical and/or instrumental noise/uncertainties  in the data ${\bf g}$, the
information about the true solution $\xi(T)$ is lost and cannot be recovered
without adding extra information about $\xi(T)$. Therefore constraints must be
applied to obtain a unique meaningful solution. All methods solving this
system to find the DEM $\xi(T)$ explicitly or implicitly add information not
present in the data to obtain the approximate solution. The
simplest, but most popular way to constraint the data is to fit a model function
$\xi(T,\alpha_i)$ with a number of free parameters $ \alpha_i$
that minimise the Eq. \ref{mproblem}. Forward fitting is highly
unsatisfactory if
the functional shape of $\xi(T)$ is {\it a priori} unknown.

As any attempt to reconstruct the DEM directly leads to substantial
noise amplification in $\xi(T)$, the broad approach to achieve a solution is to
add linear constraints to the DEM
\citep[e.g.][]{ti63,1985InvPr...1..301B,1988InvPr...4..573B,1977A&A....61..575C,
1986ipag.book.....C}. Often, so-called zero order regularization is used, which
selects the smallest
norm solution out of infinitely many possible solutions. This approach proved to
be robust for various problems and is not over restricting
\citep{1985InvPr...1..301B,1988InvPr...4..573B}.  Hence, we solve the least
square problem

\begin{equation}\label{mproblem2}
 \left\|{\bf \widetilde{K}}{\xi(T)-\bf
\widetilde{g}}\right\|^2  =  \mathrm{min}  \;
\mbox{subject to} \; \|{{{\bf L}
(\xi(T)-\xi_0(T))}}\|^2 \le \mathrm{const},
\end{equation}

\noindent with  ${\bf \widetilde{K}}=(\delta {\bf g})^{-1} {\bf K}$ and ${\bf
\widetilde{g}}=(\delta {\bf g})^{-1} {\bf g}$. This can be solved using
Lagrangian multipliers, i.e.

\begin{equation}\label{mproblem3}
\left\|{\bf \widetilde{K}}\;{\xi(T)-\bf \widetilde{g}}\right\|^2
 +\lambda\|{{{\bf L} \;(\xi(T)-\xi_0(T))}}\|^2 =\mbox{min},
\end{equation}

\noindent where ${\bf L}$ is the constraint matrix, $\lambda$ is the
regularization parameter (related to the $\chi^2$ of the solution), and
$\xi_0(T)$ is the ``guess'' solution, which will be explained in detail below.
The L2 norm is defined as a sum $\left\|{\bf x} \right\|^2={\bf x}^T{\bf x}=\sum
_{i=1}^{N} x_i^2$ over all filters or intensities. Importantly, the solution of
Eq. \ref{mproblem3} is unique and well-behaved. The formal
solution of Eq. \ref{mproblem3} $\xi_\lambda(T)$ can be simply expressed in
matrix form as a function of regularization parameter $\lambda$ but to avoid
time consuming matrix manipulations Generalized Singular Value Decomposition
GSVD \citep{1992InvPr...8..849H} is used. The GSVD of matrices
$\bf{\widetilde K}\in \mathbb{R}^{M\times N}$, $\bf{L}\in \mathbb{R}^{N\times
N}$ produces a set of singular values $\gamma_i$, $\beta_i$ and singular vectors
${\bf u}_i, {\bf v}_i, {\bf w}_i$, with $i=1,...,N$ which satisfy $\gamma_i^2 +
\beta_i^2=1$, ${\bf U^T\widetilde {K}W}=\mathrm{diag}(\gamma)$ and ${\bf V^T
LW}=\mathrm{diag}(\beta)$. These then provide the solution
\citep{1992InvPr...8..849H} to the minimization problem given in
Eq. \ref{mproblem3} as

\begin{equation}\label{eq:sol_gen}
\xi_\lambda(T) =\sum_{i=1}^ M\frac{\phi_i^2}{\phi_i^2+\lambda}\left(
\frac{({\bf g}\cdot {\bf u}_i){\bf w}_i}{\gamma_i}+\frac{\lambda \xi_0(T)}{\gamma_i^2}
\right),
\end{equation}

\noindent with $\phi_i=\gamma_i/\beta_i$. This solution weights the
contributions from various singular vectors differently, filtering out the singular
vectors with $i$, for which $\phi_i^2 < \lambda $. Hence removing un-physical
oscillatory component of the solution \citep{1988InvPr...4..573B}. Figure
\ref{fig:solution_components} shows typical behaviour of singular values and
vectors as well as the construction of the solution.

\subsection{Regularization Parameter}

To find the solution $\xi_\lambda(T)$ is to determine the
regularization parameter $\lambda$, which is done using Morozov's discrepancy
principle \citep{Morozov1967}, i.e.

\begin{equation}\label{eq:morozov}
\frac{1}{N}\left\|{\bf \widetilde K}\; \xi_\lambda(T)-{\bf \widetilde
g}\right\|^2 =\alpha,
\end{equation}

solving for $\lambda$
after substituting $\xi_\lambda(T)$ from Eq. \ref{eq:sol_gen}. Here
$\alpha$ is the regularization ``tweak'' value which effectively controls the
required $\chi^2$ of the solution in observable space. The $\alpha$ value has a
clear meaning when $\left\|{\bf \widetilde K}\; \xi_\lambda(T)-{\bf \widetilde
g}\right\|^2$ are normally distributed, with a mean of $N$ and variance of
$\sqrt{N}$. Therefore values of $\left\|{\bf \widetilde K}\; \xi_\lambda(T)-{\bf
\widetilde g}\right\|^2$ in the range $N-\sqrt{N} < N+\sqrt{N}$ are acceptable
values and $\alpha$ helps choose the exact value within this range. This also
helps to put more or less weight on the data, with $\alpha < 1$ requiring a
``better'' agreement with the data.

\subsubsection{Positively defined DEM}\label{sec:pos}
Using the method discussed in the previous section, it is possible to
select only positive solutions from the family of solutions $\xi_\lambda(T)$ by
choosing an appropriate $\lambda$. The method based on Morozov's
discrepancy principle chooses the parameter $\lambda $ that gives the DEM
solution $\xi_\lambda(T)$ with the desired $\alpha$ in Eq. \ref{eq:morozov}.
The intrinsic DEM from the Sun should be positive but the regularization method
provides no guarantee of a positive solution. Although this appears to be a
problem with this implementation of the regularization method, one should
remember that the DEM derived from observations need not be positive given
the often poorly known uncertainties, response functions and the possibility of
background subtraction.

A positive DEM solution can be achieved with an additional criterion to
the choice of regularization parameter. That is we take the regularization
parameter $\lambda$ that provides the smallest $\left\|{\bf \widetilde K}\;
\xi_\lambda(T)-{\bf \widetilde g}\right\|^2 -\alpha N$ and $\xi_\lambda(T)>0$.
This approach has the advantage of maintaining the linear calculation of the
solution unlike those that try to implement the positivity constraint in Eq.
\ref{eq:sol_gen} directly, producing non-linear or iterative solutions
\citep[e.g][]{1997InvPr..13..441P,1999InvPr..15..615D}. As $\alpha$ is the
$\chi^2$ of the regularized solution, the ability to recover a positive solution
strongly depends on the error estimates on the input data and knowledge of ${\bf
K}$. If the error used is too small the $\chi^2$ of a positive solution can be
erroneously high and in general, the positivity constraint produces a larger
$\chi^2$, behaviour previously demonstrated by \citet{1981AcOpt..28.1635B}.

\subsection{Initial Guess Solution}\label{sec:guess}

The standard mode of operation of our regularization algorithm requires
no initial guess solution, i.e. $\xi_0(T)=0$. However, $\xi_0(T)$ is used in the
calculation of the constraint matrix (either for the higher order constraints or in
the constraint weighting, see \S\ref{sec:conmat}). To avoid this problem we run
our regularization algorithm (solving Eq. \ref{eq:morozov} and then Eq.
\ref{eq:sol_gen}) twice. On the first run the guess solution is $\xi_0(T)=0$,
the
constraint matrix is the identity matrix ${\bf L}={\bf I}$ and we find a weakly
regularized solution with $\alpha=10$. The regularized solution found can then
be used as the initial guess for the second run ($\xi_0(T)=\xi_R(T)$) and in
calculating the chosen constraint matrix (as discussed in \S\ref{sec:conmat}).
For this second run a stronger regularization is used $\alpha=1$ to find the final
solution, then the associated errors in the DEM and temperatures are
calculated, see \S\ref{sec:errors}.

An alternative approach can be taken when working with high resolution
spectroscopic line data as we can use the minimum of the EM loci curves as the
 initial guess solution. The EM loci curve for each spectral line is the ratio
of the line intensity to contribution function $\approx g_i/K_i$. As this estimates
the EM based on the isothermal temperature, it provides the upper limit to the
DEM as a function of temperature. Multiple spectral lines across a wide range
of temperatures provides a strong constraint to the DEM space and a useful
initial guess solution. When selected it is automatically calculated
within our code and as we are starting with a non-zero guess solution, this
approach only requires one run of the algorithm. These two approaches to the
guess solution will be demonstrated for Hinode/EIS data, simulated from a
variety of model DEMs, in \S3.2 and \S4.2. For broadband data (such as those
from SDO/AIA), only the $\xi_0(T)=0$ approach is used as the minimum of the
EM loci curves provides a poor guess solution, especially when the DEM is not
isothermal.

\subsection{Constraint matrix}\label{sec:conmat}
There is a number of different choices for ${\bf L}$ and here we consider only
linear constraints.  Physically, the quantity $\int n_e^2 dh= \int \xi(T)
dT$ is the total number of electrons along the line of sight, so $\|\xi(T)\|^2\le
\mathrm{constant}$, similar to the X-ray case \citep{2003ApJ...595L.127P}.  This
corresponds to the constraint matrix of ${\bf L}_0\propto {\bf I}$, a zeroth-order
constraint. So, applying zero order regularization, we find DEM with the smallest
amount of plasma required to explain the observational data.  The source
averaged $\xi(T)$ results from a combination of heating, cooling and and the
physics of heat transport in the radiating source. Therefore, when $\xi(T)$ is the
solution of some differential equation, one can expect that $\xi(T)$ should be
differentiable or equivalently the constraint matrix ${\bf L}$ is the first order or
second order derivative (${\bf L}_1\sim {\bf D}^1$ or ${\bf L}_2\sim {\bf D}^2$),
again similar to the  X-ray case \citep{2004SoPh..225..293K}. Hence higher order
regularization constraints select solutions with the smallest variations in
temperature and are more restrictive then zeroth-order solutions. The
more restrictive methods put more weight on the a priori constraints rather than
the analysed data sets, which could be advantageous for poorly measured data.

\subsection{Error and Temperature Resolution of DEM }\label{sec:errors}

Suppose the true DEM solution $\xi_{true}(T)$ is given. Then we can write

\begin{equation}\label{eq:g2}
{\bf g} = {\bf K} {\xi}_\mathrm{true}(T)+ \delta {\bf g}.
\end{equation}

\noindent  Any regularized solution to a linear problem can be viewed as the
replacement of the generalized inverse ${\bf K}^{+}$  with the regularized
inverse ${\bf R}_\lambda$ (with $\lim_{\lambda \to 0} {\bf R}_\lambda={\bf
K}^+$), so that our regularized solution is $\xi_R(T) = {\bf R}_\lambda {\bf
g}$. Indeed for $\xi_0(T)=0$, Eq. \ref{eq:g2} can straightforwardly
re-written in such form with ${\bf R}_\lambda$ expressed via known GSVD vectors
and values and exists for all linear methods \citep{1988InvPr...4..573B}.
On the other hand, the true solution to Eq. \ref{eq:g2}
can be formally written ${\bf K^{+}g}$. Therefore, to estimate the error we find
the difference between the true solution and our solution

\begin{equation}\label{eq3}
\delta \xi(T) = \xi_R(T)- \xi _\mathrm{true}(T) =
 ({\bf R}_\lambda{\bf K} -{\bf I}) \xi _\mathrm{true}(T)
+{\bf R}_\lambda \delta {\bf g}
\end{equation}

\noindent Eq. \ref{eq3} shows the important result that the error comes
from two parts: the last term gives us that the noise propagation (vertical
errors) and the first term gives the temperature resolution (horizontal errors).
This equation presents the method independent definition of both the horizontal
and vertical resolutions . While the noise propagation is normally accounted by
the DEM methods, the resulting temperature resolution is often not
considered.

\begin{figure*}\centering
\includegraphics[width=160mm]{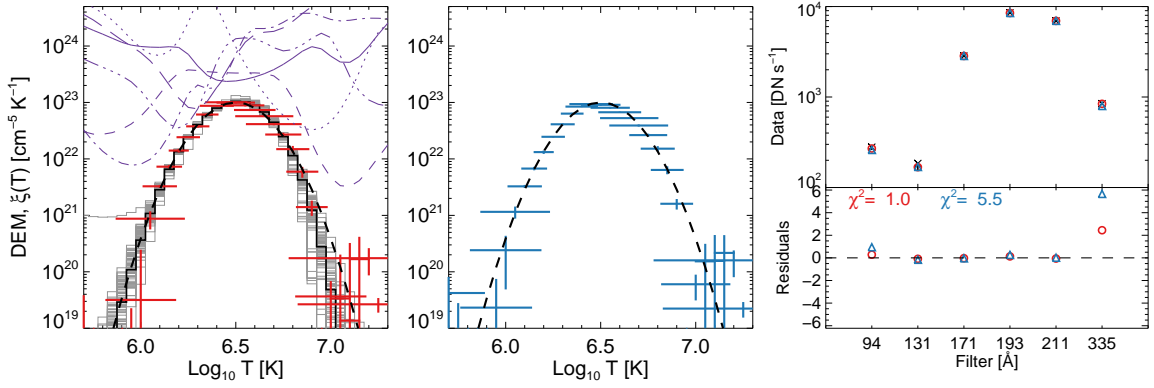}
\caption{\label{fig:aia_exae} Results of DEM reconstructions from
simulated SDO/AIA data of a single Gaussian model. Shown is the zeroth-order
regularized solution (left panel, red error bars) and with an additional
positivity
constraint (middle panel, blue error bars). The resulting simulated SDO/AIA data
for each filter and the residuals of the regularized solutions are shown in the
upper and lower right panels. Also plotted is the solution from
\texttt{xrt\_dem\_iterative2.pro} (grey histogram, left panel) and the EM loci
curves (purple lines, left panel).}
\end{figure*}

The error on the DEM $\delta \xi(T)$  (the vertical error ${\bf
R_\lambda}\delta {\bf g}$) is calculated using the standard Monte Carlo
approach \citep{1992nrca.book.....P,2003ApJ...595L.127P,2004SoPh..225..293K,
2006SoPh..237...61P} with multiple random realisations of ${\bf g}$ within the
noise range $\delta {\bf g}$. Then the one sigma spread of the regularized
solutions from these realisations provides the measure of the uncertainty on the
DEM. This is possible due to the linear nature of Eq. \ref{eq3}. In general,
the
exact statistics of the errors is needed but as this distribution is unknown, we
assume a Gaussian distribution of standard deviation $\delta {\bf g}$ for each
filter. Then the the vertical uncertainty on the DEM are calculated as the
standard deviation using 300 MC Gaussian noise realisations. However, as the
probability distribution of $\delta {\bf g}$ does not need to be Gaussian, the
resulting DEM uncertainties could also be non-Gaussian.

The temperature resolution (horizontal error) of any linear inversion
method -- or temperature bias of the solution -- is how much the product ${\bf
R}_\lambda{\bf K}$ differs from the  identity matrix ${\bf I}$. When the
regularized inverse is similar to inverse of the kernel matrix ${\bf
R}_\lambda\simeq {\bf K}^+$, the temperature resolution does not degrade giving
${\bf R}_\lambda{\bf K} \simeq {\bf I}$. However, for the ill-conditioned
problem of DEM determination, ${\bf R}_\lambda{\bf K} $ is not identity matrix,
but has a finite spread. The temperature resolution is then simply the FWHM of
${\bf R}_\lambda{\bf K} $ for a given temperature bin. This represents the
temperature bias measure or the smallest temperature difference which can be
meaningfully distinguished in the solution. Conveniently, ${\bf {R_\lambda}
{K}}$ is easy to calculate from the singular values and vectors obtained in GSVD
decomposition of ${\bf \widetilde K}$ and ${\bf L}$ used to find the regularized
solution. Namely, ${\bf R_\lambda K}={\bf WYW^{-1}}$ where the column vector
${\bf w}_i$ forms matrix ${\bf W}$ and ${\bf Y}$ is a diagonal matrix with the
elements constructed with singular values ${\bf
Y}_{ii}=\gamma_i^2/(\gamma_i^2+\lambda \beta_i^2)$ \citep{2004SoPh..225..293K}.

\begin{figure}\centering
\includegraphics[width=85mm]{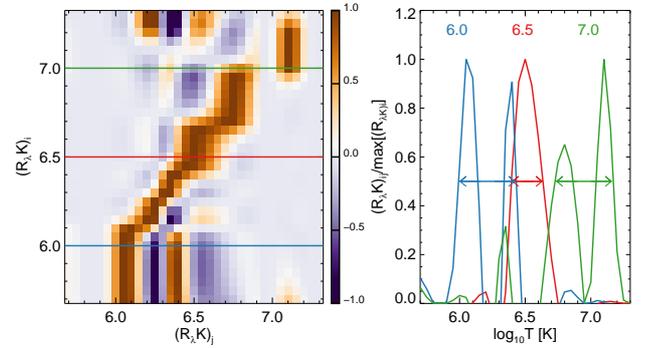}
\caption{\label{fig:aia_exrr}
(Left) ${\bf R_\lambda}{\bf K}$ matrix with each row providing the temperature
resolution information for each temperature $T_j$. The highlighted rows are
plotted in the right panel also showing the FWHM estimate used to calculate
the temperature resolution.}
\end{figure}

When ${\bf R}_\lambda{\bf K}={\bf I}$ the response is diagonal or
impulse-like for a given temperature, and therefore the true DEM
${\xi}_\mathrm{true}(T)$ is not distorted.  In any practical situation, ${\bf
R}_\lambda{\bf K}$ is not diagonal but has off-diagonal elements (an example is
shown in Figure \ref{fig:aia_exrr} of a simulated DEM discussed in detail in
\S\ref{sec:aia_gauss}). The off-diagonal terms are a spread about a peak value
at the diagonal (the red example in Figure \ref{fig:aia_exrr}), so for each
temperature $T_j$ the row of $({\bf R_\lambda K})_i$ has a finite width. The
FWHM of this spread is then taken as the temperature resolution for that
particular temperature. When the off-diagonal terms dominate (the blue and green
examples in Figure \ref{fig:aia_exrr}), the FWHM can still be used to indicate
the poor temperature resolution. It should be noted that  ${\bf R}_\lambda{\bf
K}$ depends on the errors $\delta {\bf g}$, so larger errors results in a
poorer temperature resolution, e.g. wider row of the matrix ${\bf R}_\lambda{\bf
K}$.
The dependency on $\delta {\bf g}$ comes via the regularization parameter
$\lambda$ given by Eq. \ref{eq:morozov}. This definition of temperature
resolution is more conservative as it does not assume a form of DEM as in
\citet{2009ASPC..415...32W}.

\begin{figure}\centering
\includegraphics[width=95mm]{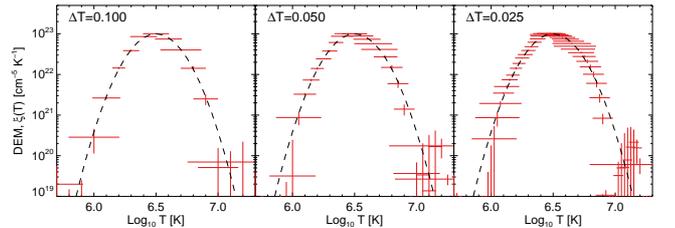}
\caption{\label{fig:aia_exnt}Effect of changing the temperature bin size for
the single Gaussian model DEM, shown in Figure \ref{fig:aia_exae}.}
\end{figure}

\subsection{Temperature response and instrument uncertainties}
\label{sec:terr}

One of complicating factors is that the kernel (the matrix ${\bf K}$)  of the
integral equation to be inverted is only known to a limited degree. The errors
come from the calibration uncertainty of the instrument itself and the
uncertainty in the dominant spectral contribution of each bandpass
\citep[e.g.][]{2011ApJ...732...81A,2011A&A...525A.137O}. Therefore, the
temperature response has an uncertainty $ {\bf \delta K}$ and the linear
problem, Eq. \ref{eq:g2}, becomes

\begin{equation}\label{eq:deltaK}
{\bf g} = ({\bf K} + {\bf \delta K}){\xi}_\mathrm{true}(T)+ \delta {\bf g} \; .
\end{equation}

\noindent This translates into an additional uncertainty for $\xi(T)$ compared
to Eq. \ref{eq3}

\begin{equation}\label{eq:new1}
\delta \xi(T) =
 ({\bf R}_\lambda{\bf K} -{\bf I}) \xi _\mathrm{true}(T)
+{\bf R}_\lambda{\bf K}\;  \delta {\bf g} + {\bf R}_\lambda{\bf \delta K}\; \xi
_\mathrm{true}(T)
\end{equation}

\noindent When the uncertainty $ {\bf \delta K}$ is dominated by a systematic
error in the intensity of the lines measured then it is not temperature
dependent but a constant scaling factor per filer or line (i.e. rows of ${\bf
K}$). Since the shape of the response or contribution function does not change
as a function of temperature then this is identical to the introduction of an
additional error to $\delta {\bf g}$ with Eq. \ref{eq:new1} reducing to Eq.
\ref{eq3}. This will be investigated further in \S\ref{sec:eis_gauss} and
\S\ref{sec:eis_chi} with reference to Hinode/EIS.

\section{Simulated Data: Gaussian Model}\label{sec:mod_gauss}

We test the regularization method on simulated SDO/AIA (\S\ref{sec:aia_gauss})
and Hinode/EIS data (\S\ref{sec:eis_gauss}) of Gaussian model DEMs. These
have the form

\begin{equation}\label{eq:modg}
\xi(T_j)=\frac{N_0}{\sqrt{2\pi}\sigma_T}\exp\left[\frac{-(\log{T_j}-\log{T_0}
)^2 } { 2\sigma_T^2 } \right]
 \end{equation}

\noindent where $\log{T_0}$ is the centroid temperature, $\sigma_T$ is the
standard deviation and $N_0=\int \xi(T) dT$.  We consider model DEMs of one
to three Gaussian components. Note that as we calculate everything using
$\log{T}$ instead of $T$, a conversion factor of $T \ln{(10)} d\log{T}$ is
required in Eqs. \ref{eq:dem_int} and \ref{eq1}.

\subsection{SDO/AIA Simulated Data}\label{sec:aia_gauss}

To simulate SDO/AIA data we take the model DEM from Eq. \ref{eq:modg}
and calculate the expected observable signal in each of AIA's six coronal
filters $g_i$ using the response functions $K_i$ \citep{Boerner}, shown in the
top panel of Figure \ref{fig:tr}. From this we calculate the associated error
$\delta g_i$ using the readout noise and photon counting statistics (correcting
from DN s$^{-1}$ px$^{-1}$ to photons via the electron and photon gains). Then
Gaussian noise within these $\delta  g_i$ is added to $g_i$. Our simulated
observables ${\bf g}, \delta {\bf g}$ and the response functions ${\bf K}$ are
the inputs to the regularization algorithm. As described in \S \ref{sec:reg} we
run the regularization twice: on the first run we use $\alpha=10, {\bf L}={\bf
I}, \xi_0(T)=0$ which provides a guess solution that can be used to calculate
the desired constraint matrix ${\bf L_0, L_1, L_2}$ and final regularized
solution with $\alpha=1$.

We first consider a single Gaussian model of $N_0=3.76\times10^{22}$
cm$^{-5}$, $\sigma_T=0.15$ and $\log{T_0}=6.5$, shown in Figure
\ref{fig:aia_exae}. Here the regularized DEM (red error bars) was found using
${\bf L_0}$ (zeroth-order constraint) and well matches the original model DEM
(black dashed line). In observable space (right panel in Figure
\ref{fig:aia_exae})
the residuals between the model and regularized solution are small with $\chi^2
\approx 1.0$ close to the desired value set by $\alpha=1$. Here $\chi^2$ is
taken as the sum of the square of the residuals divided by $N$ to match the
version in Eq. \ref{eq:morozov}. The ${\bf R_\lambda K}$ matrix for this
regularization (Figure \ref{fig:aia_exrr}) is diagonal over approximately
$\log{T}=6.1 - 6.9$. The horizontal error bars (temperature resolution) at each
temperature bin is taken as the standard deviation of the FWHM of the rows of
${\bf R_\lambda K}$. Outside this range the matrix is clearly not diagonal,
producing large horizontal errors, so the regularized DEM is not reliable at
these temperatures.

At the lowest temperatures (log $T\le 6.0$) some negative DEM values
are found, not shown on the log-scale, but this is in a temperature range where
the DEM contribution is minuscule (over 4 orders of magnitude smaller than the
peak) and the temperature response is very weak. At the highest temperature
(log$T\ge 7.0$) we have positive regularized DEM but with large horizontal and
vertical errors again due to being in a range where the DEM component is very
small.  In the middle panel of Figure  \ref{fig:aia_exae} we show the
regularized
solution (blue error bars) in which the regularization parameter has been chosen
to minimise $\chi^2$ to the desired value whilst forcing a positive DEM. The
result is a very slightly different DEM, particularly below log$T \approx 6.1$,
and
is well within the error bounds shown for both regularized solutions. As
expected the $\chi^2$ of the solution is higher ($5.5$ instead of the desired
$1$) but within the DEM error bounds and so both are valid solutions.

\begin{figure*}\centering
\includegraphics[width=160mm]{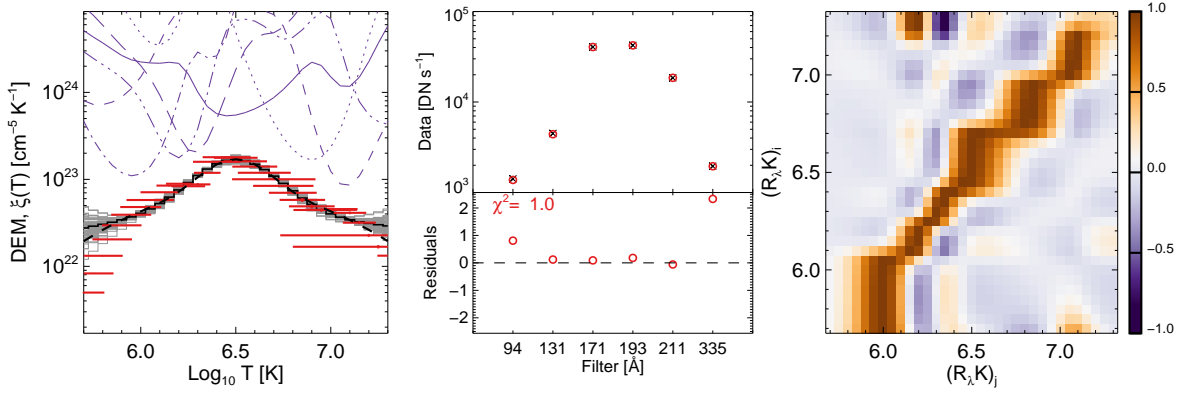}
\caption{\label{fig:aia_example2} Example DEM from zeroth-order
regularization of simulated SDO/AIA from a two Gaussian model. (Left) DEM
space, (middle) observable space and residuals (right) ${\bf R_\lambda K}$ used
to calculate the temperature resolution. Also plotted is the solution from
\texttt{xrt\_dem\_iterative2.pro} (grey histogram, left panel) and the EM loci
curves (purple lines, left panel). Note that the solution here is positive
without
the additional constraint.}
\end{figure*}

\begin{figure}\centering
\includegraphics[width=80mm]{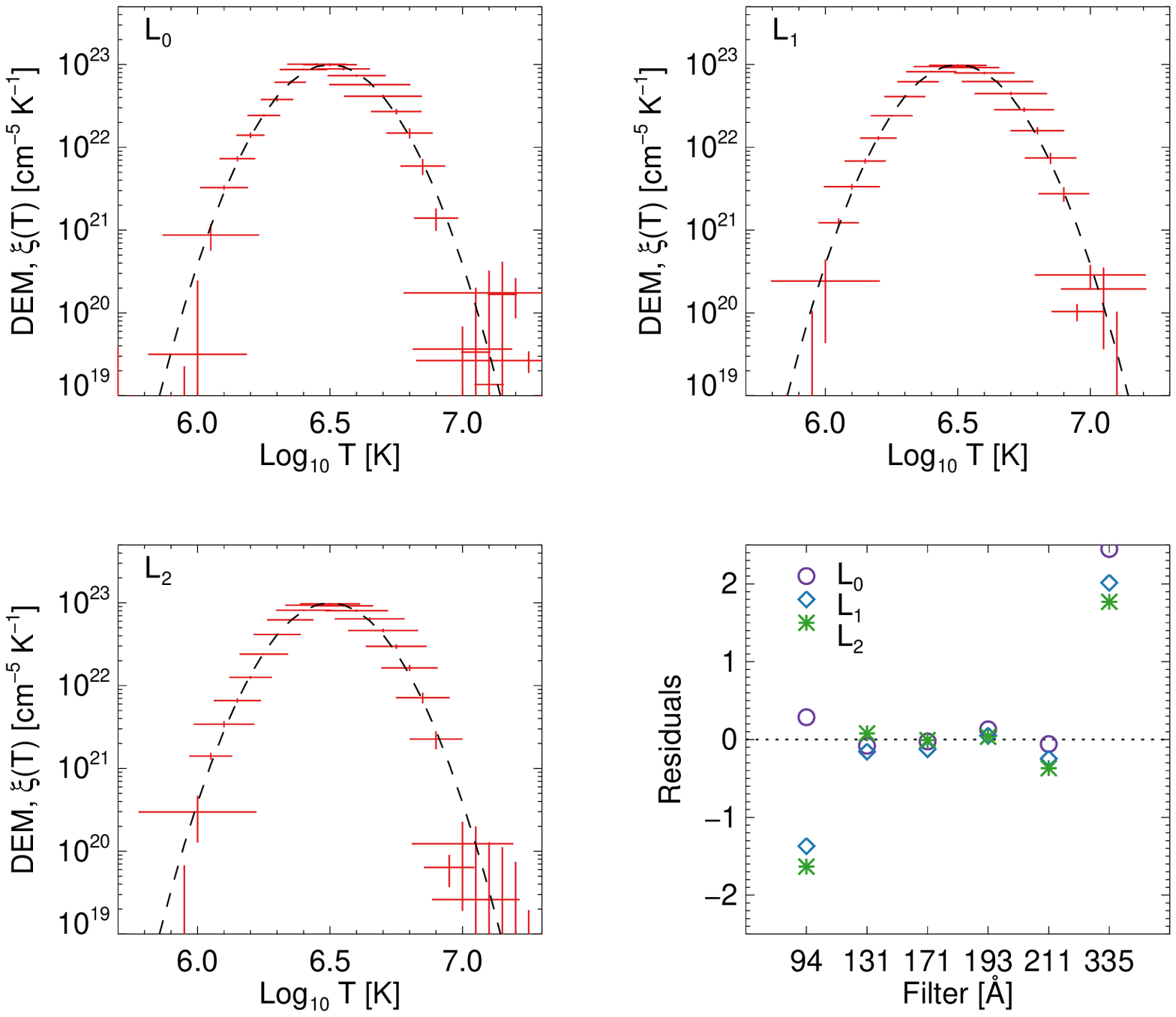}\\
\includegraphics[width=80mm]{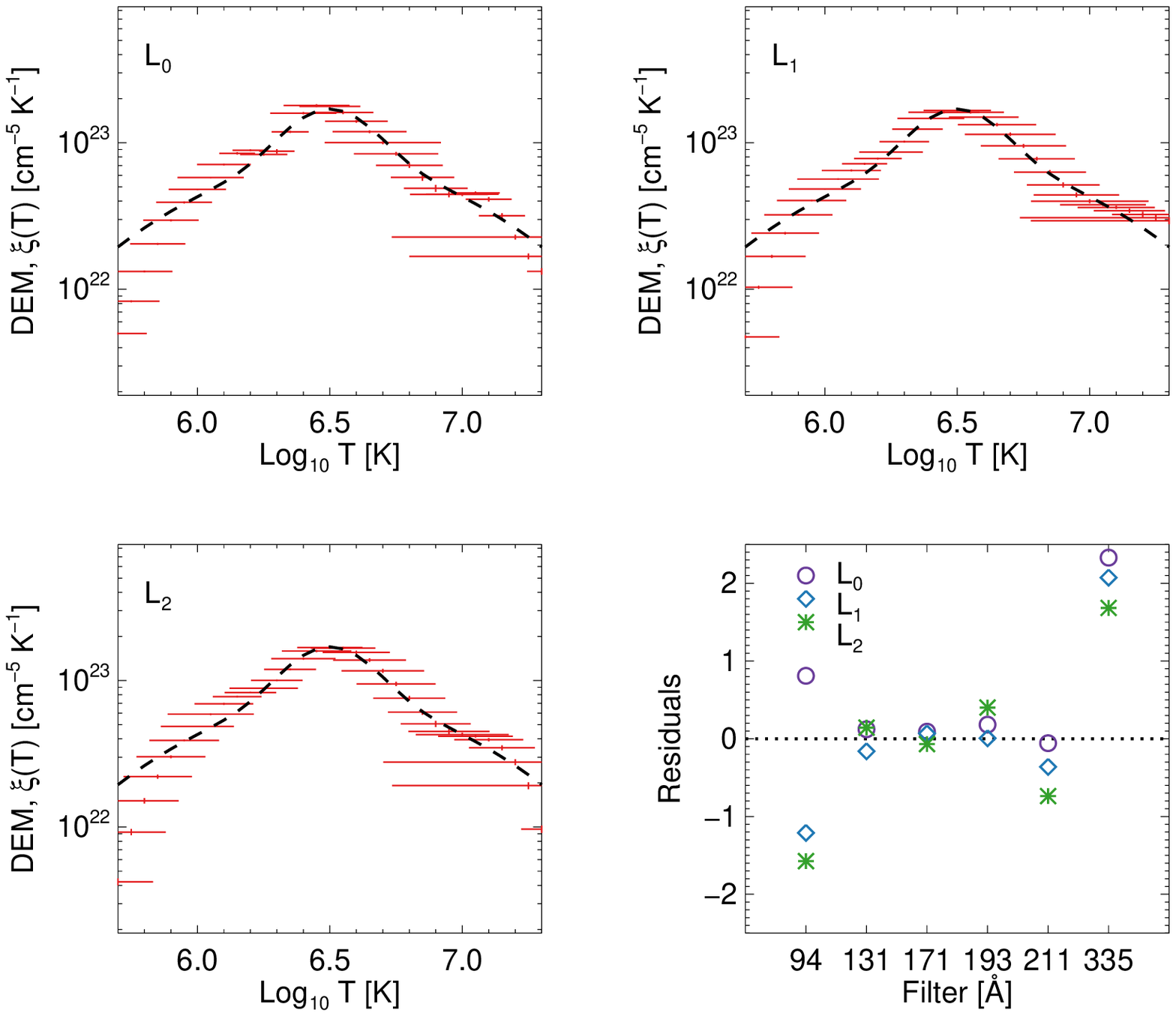}
\caption{\label{fig:aia_exord}Testing the effect of changing the constraint
matrix order (${\bf L_0, L_1, L_2}$, zeroth, first and second order
respectively).
The top four panels show the different recovered DEM and resulting residuals
for the single Gaussian model shown in Figure \ref{fig:aia_exae}. The bottom
four panels shows the same for the broad double Gaussian model shown in
Figure \ref{fig:aia_example2}.}
\end{figure}

For comparison we have calculated the DEM solution using the iterative forward
fitting routine \texttt{xrt\_dem\_iterative2.pro} (black histogram in left panel
Figure \ref{fig:aia_exae}) with 75 Monte Carlo (MC) realisations of the solution
within the observable error bound $g_i \pm \delta g_i$ (grey histograms). This
also agrees well with the model DEM and also shows a larger spread in DEM
solution in the temperature ranges where the regularized solution has large
errors. Again this is due to the minor contribution to the DEM (about four
orders of magnitude smaller than the peak value) in these temperature ranges.
Also for comparison we have plotted the EM loci curves ($g_i/(K_i T d\ln{T})$)
which do not intersect at a single point as this is not an isothermal model DEM.
The DEM solutions are below the EM loci curves, as expected, since these
estimate the upper limit of the emission.

The error bars shown for the regularized solution are not independent of each
other and map out an error boundary region. This can be seen in Figure
\ref{fig:aia_exnt} where we show the regularized DEM using a variety of
temperature binsizes. To achieve this we need to interpolate the temperature
response functions (top panel Figure \ref{fig:tr}) from the original binning to
our chosen temperature binsize. Increasing the number of temperature bins does
not change the shape of the regularized DEM but produces a clearer definition of
the error bound, with the overlapping error bars indicating that the nearby DEM
bins are clearly not independent. This does however slow down the computation of
the regularization process, although at worst it still only took a few seconds
to compute. Note again that the vertical errors are taken as the variance of 300
regularized solutions found from random realisations within $g_i \pm \delta g_i$
and so we are explicitly assuming a simple Gaussian spread of vertical error in
the regularized solutions. In reality the distribution of these errors will be
more complicated.

A second example is shown in Figure \ref{fig:aia_example2} where an additional
broad Gaussian component ($N_0=8.77\times10^{22}$ cm$^{-5}$,
$\sigma_T=0.5$ and $\log{T_0}=6.5$) has been added to the model DEM shown
in Figure \ref{fig:aia_exae}. Immediately it is clear that the regularization
produces a ${\bf R_\lambda K}$ that is diagonal over a wider temperature
range, indicating that the narrow temperature range found for the single
Gaussian model (Figure \ref{fig:aia_exae}) was mostly due to the DEM model
dominating over a small temperature range. For the broader DEM model there
are still deviations from a diagonal ${\bf R_\lambda K}$ at the ends of the
temperature range chosen but this is due to the limited response of AIA at these
temperatures (see Figure \ref{fig:tr}). This time only the solution without
a positivity constraint is shown since the regularization parameter for
$\chi^2=1$ provides a positive solution. For comparison the forward fitted
solutions from \texttt{xrt\_dem\_iterative2.pro} (grey histograms) are shown
and again match the regularized solution (and model DEM) over the majority of
the temperature range but show a wider vertical spread at the smallest and
largest temperatures. This consistent increase in error in the DEM solution from
both methods suggests the poor response at these temperature extremes is the
source of this uncertainty. The EM loci curves are also shown but again do not
intersect at a single point as this is a broad DEM.

In both examples shown so far only the zeroth-order constraint has been used
and in Figure \ref{fig:aia_exord} we show the resulting regularized DEM for
these cases for higher order constraint matrices. For the single Gaussian the
higher order constraint removes the cluster of ``noisy'' data points at high
temperature but also broadens the temperature resolution at around
$\log{T_0}=6.2-6.4$. This shows that the constraint is changing the balance
between these two temperature ranges and in this case the lower order
constraint is preferable as it produces a better temperature resolution where
the
DEM is dominant. For the two Gaussian model DEM the higher order constraints
again produce a broader temperature resolution which helps the regularized
solution match the model  better at low temperatures. Therefore in this case the
higher order constraint is marginally preferable. As expected, in both
cases the increase in order of the constraint matrix increases the ``smoothing''
of the regularized solution. In the subsequent analysis present here the zeroth
order $L_0$ constraint matrix is used throughout as it is generally sufficient
to
recover the expected DEM.

\begin{figure*}\centering
\includegraphics[width=140mm]{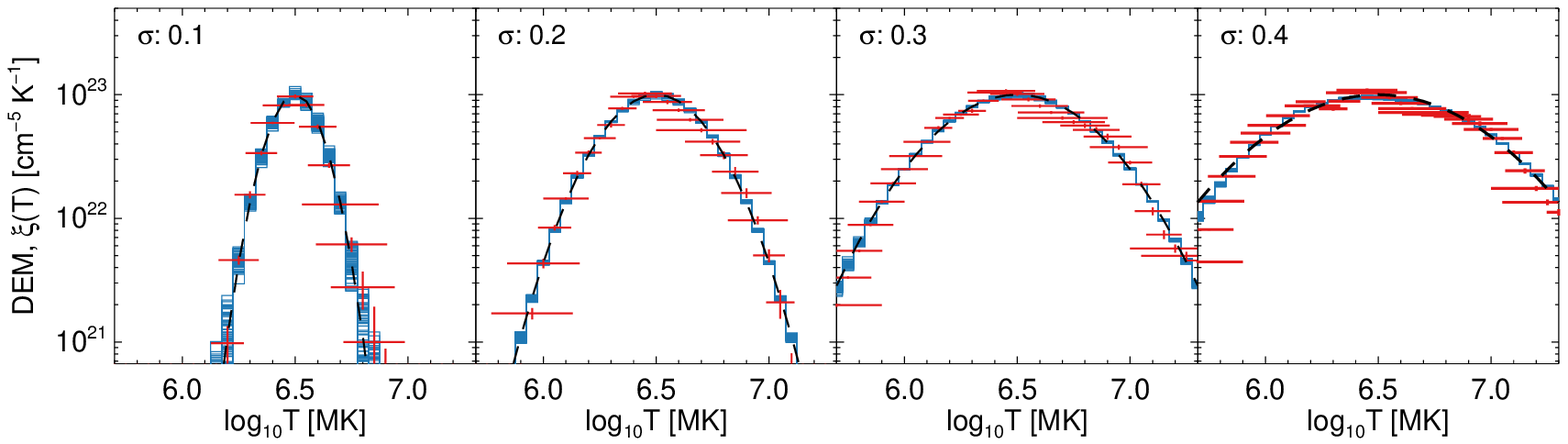}\\
\includegraphics[width=140mm]{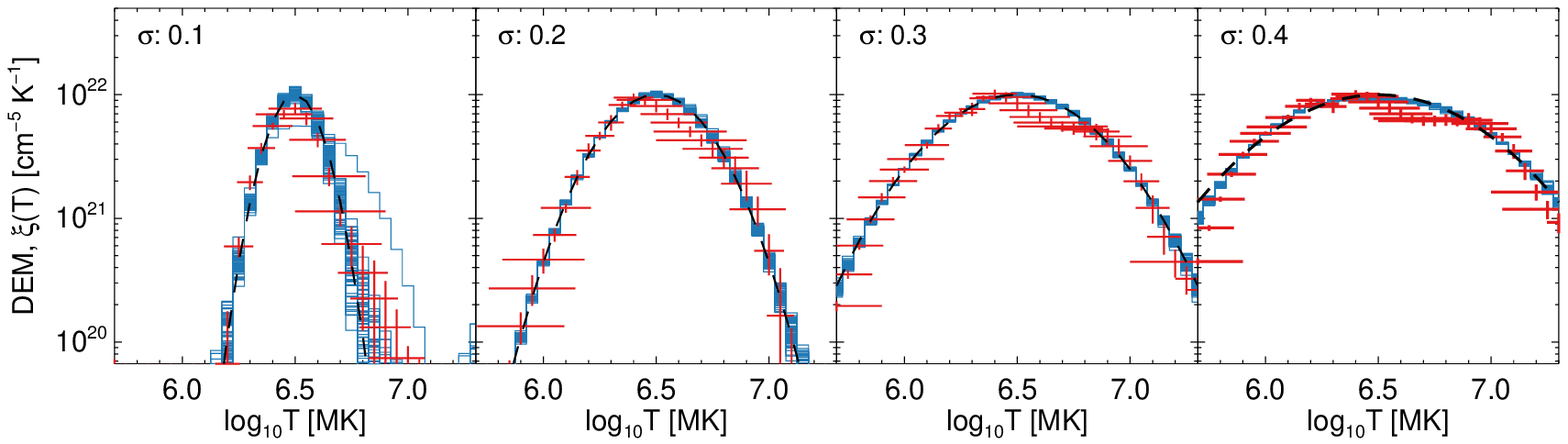}
\caption{\label{fig:aia_msigma}DEMs recovered using regularized
inversion (red error bars) and  \texttt{xrt\_dem\_iterative2.pro} (blue
histograms, with 75 MC realisations) from simulated SDO/AIA data of Gaussian
models (black dashed lines) of differing $\sigma$ (increasing left to right) and
normalisation magnitude ($10^{23}$ top row, $10^{22}$ bottom row).}
\end{figure*}

\begin{figure*}\centering
\includegraphics[width=140mm]{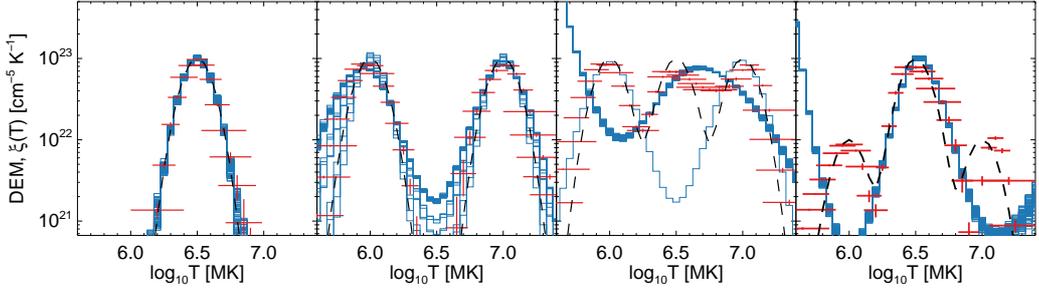}
\caption{\label{fig:aia_mg}DEMs recovered using regularized inversion
(red error bars) and  \texttt{xrt\_dem\_iterative2.pro} (blue histograms, with
75
MC realisations) from simulated SDO/AIA data of models (black dashed lines)
with a differing number of Gaussian components (left to right).}
\end{figure*}

We now consider the ability of the regularized inversion to recover
single
Gaussian model DEMs but with different widths and magnitudes, shown in
Figure \ref{fig:aia_msigma}. Here we do not use the positivity constraint since
for narrow DEMs the possible negative regions are from temperatures where the
contribution is tiny and consistent (within the errors) with zero and for broad
DEMs the are no negative values recovered. As we set $\alpha=1$ the $\chi^2$
of the regularized solution is also approximately 1. Firstly, we increase the
width
of the Gaussian from the minimum expected of $\sigma_T=0.1$, the designed
achievable temperature resolution of SDO/AIA \citep{2010ApJ...708.1238J}, to
$\sigma_T=0.4$, using a normalisation magnitude of
$N_0/(\sqrt{2\pi}\sigma_T)=10^{23}$cm$^{-5}$K$^{-1}$ in Eq. \ref{eq:modg},
shown in the top panels of Figure \ref{fig:aia_msigma}. The regularized
inversion method is able to recover the model DEM at the limit of SDO/AIA's
temperature resolution. For the widest model DEMs there is some slight
deviation but it is consistent within the error bars. Again we compare the
regularization method to \texttt{xrt\_dem\_iterative2.pro} (blue histograms) and
find similar solutions, with both methods deviating from the model DEM in
similar temperature ranges, for instance below log$T\approx 5.9$ in the
$\sigma_T=0.4$ model. We repeat the exercise but with model DEMs an order
of magnitude smaller, $N_0/(\sqrt{2\pi}\sigma_T)=10^{22}$cm$^{-5}$K$^{-1}$,
shown in the bottom panels of Figure \ref{fig:aia_msigma}. With the reduced
signal to noise both methods produce poorer DEM solutions, particularly for the
narrowest Gaussian model showing the expected degradation of the resolution
recoverable with increasing noise. In this case one of the
\texttt{xrt\_dem\_iterative2.pro} solutions deviates greatly above log$T\approx
6.7$. For the broader DEMs the iterative forward fitting method produces DEMs
closer to the model than the regularized inversion, though the later is mostly
consistent within the indicated error bars. For the regularized inversion a
better
or ``smoother'' solution could be archived by using a higher order constraint
matrix however with real data there is no prior knowledge to the form of the
emission and therefore no indication if such a constraint is actually better.

\begin{figure*}\centering
\includegraphics[width=140mm]{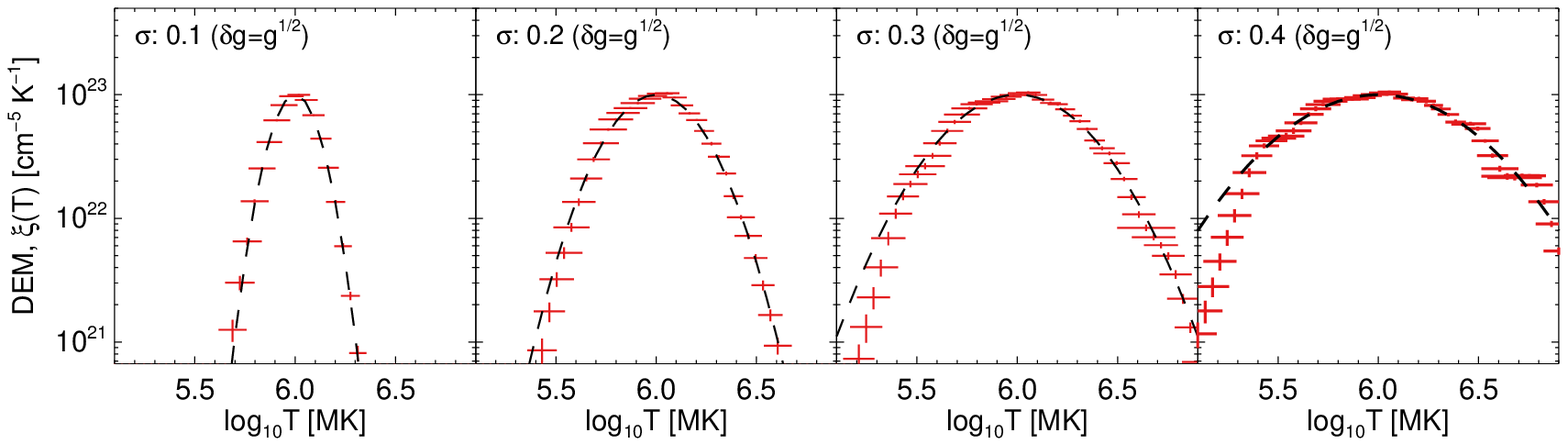}\\
\includegraphics[width=140mm]{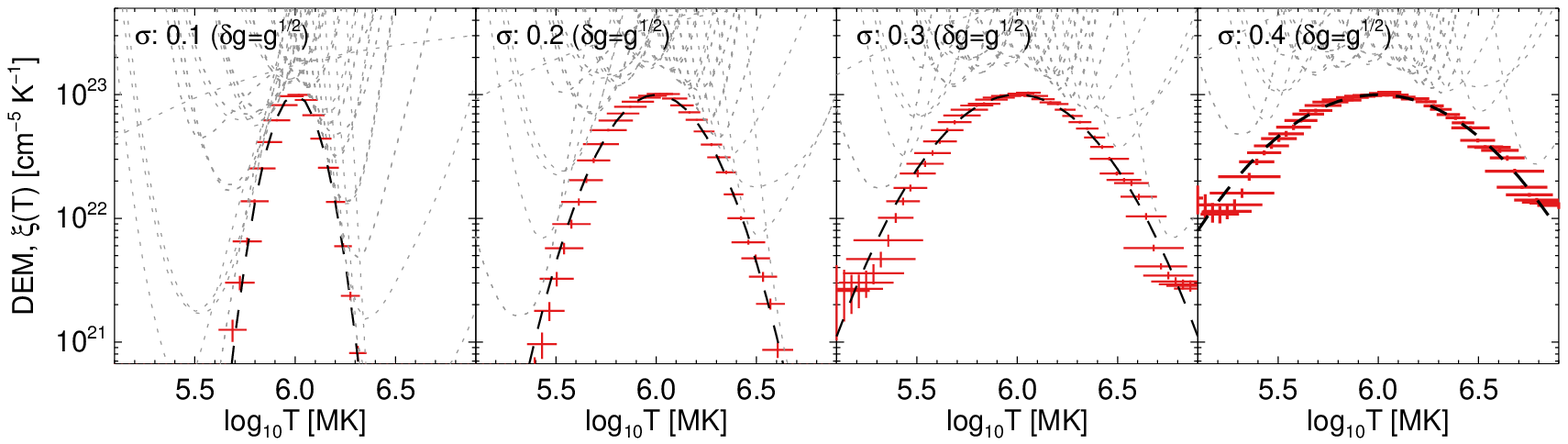}
\caption{\label{fig:eis_poi}DEMs recovered using regularized inversion
(red error bars) from simulated Hinode/EIS data of Gaussian models (black
dashed lines) of differing $\sigma$ (increasing left to right). Here the error
in
the line intensity is taken as the Poisson noise, which range between 0.2 to
11\% of the line intensity. In the bottom row the minimum of the EM loci curves
(grey dotted lines) has been used as the guess solution $\xi_0(T)$.}
\end{figure*}

\begin{figure*}\centering
\includegraphics[width=140mm]{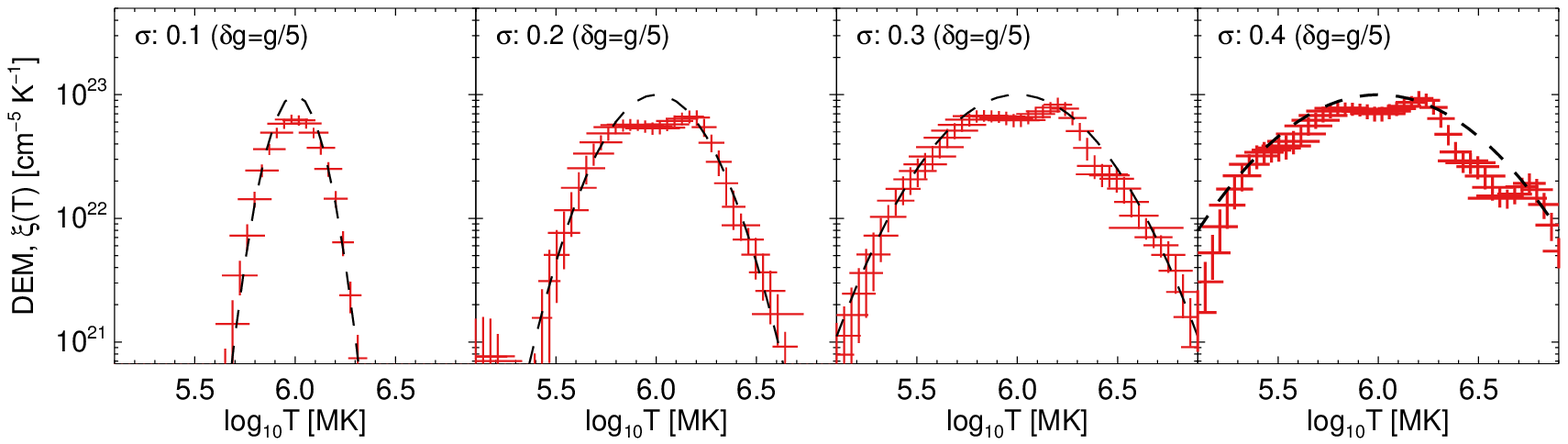}\\
\includegraphics[width=140mm]{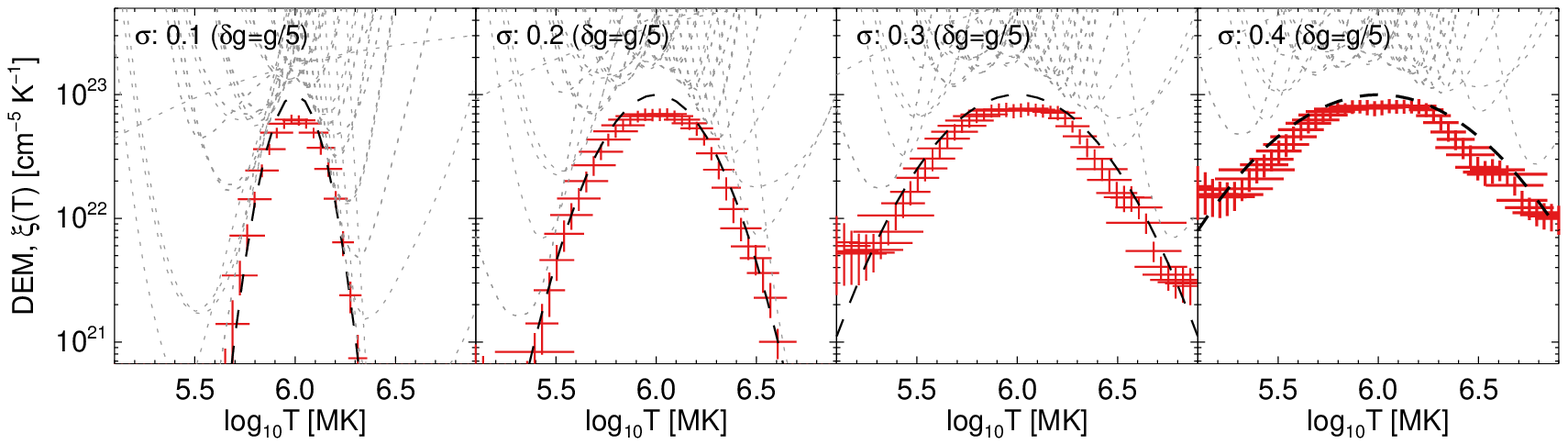}
\caption{\label{fig:eis_per}DEMs recovered using regularized inversion
(red error bars) from simulated Hinode/EIS data of Gaussian models (black
dashed lines) of differing $\sigma$ (increasing left to right). Here the error
is
taken as 20\% of the line intensity. In the bottom row the minimum of the EM
loci curves (grey dotted lines) has been used as the guess solution
$\xi_0(T)$.}
\end{figure*}

In Figure \ref{fig:aia_mg} we consider model DEMs constructed of
several Gaussian components of width at the temperature resolution of
SDO/AIA, $\sigma_T\approx 0.1$. The first panel shows the same single
Gaussian shown in Figure \ref{fig:aia_msigma} for comparison. With two
Gaussian components with centroid temperatures of log$T=6.0,7.0$ (second
panel in Figure \ref{fig:aia_mg}) the regularized solution closely matches the
source model. The forward fitting approach (\texttt{xrt\_dem\_iterative2.pro})
does get the general shape correct but produces a tall spread of solutions at
the
extremes of the temperature range chosen and in between the two Gaussian
components. The former will be due to the reduced SDO/AIA response in this
range but for the discrepancy about log$T=6.5$ this is not the case. The
situation is even worse for \texttt{xrt\_dem\_iterative2.pro} when three
Gaussian components are considered (third panel in Figure \ref{fig:aia_mg},
with centroid temperatures of log$T=6.0,6.5,7.0$) with the it producing a DEM
solution completely different from the model. In one of the MC realisations the
forward fitting method does recover two of the Gaussian components but still
fails to recover  the middle one. The regularized inversion method also has
problems in recovering the model DEM but performs considerably better than
\texttt{xrt\_dem\_iterative2.pro}, producing a DEM with three distinct
components. Only between log$T=6.4-6.8$ does it deviate from the model,
failing to match the second Gaussian peak and subsequent minima. In the final
panel of Figure \ref{fig:aia_mg} we again consider three Gaussian components
but this time with those of centroid temperatures log$T=6.0,7.0$ an order of
magnitude smaller than the central component. This time
\texttt{xrt\_dem\_iterative2.pro} recovers the central component but completely
missing the two smaller ones. Again the regularized inversion performs far
better recovering three distinct components though slightly overestimates the
centroid position of the hottest component.

\subsection{Hinode/EIS Simulated Data}\label{sec:eis_gauss}

Spectroscopic observations, such as those from the EUV Imaging Spectrometer
(EIS) on Hinode, can potentially recover the DEM better than
broadband multi-filter
observations given the significant number of temperature sensitive lines
available. The resulting observed line intensities and calculated contribution
functions can be easily used with our regularization method as it only requires
an observable, associated error and response as a function of temperature for a
number of filters or lines. Here we use the atlas of EUV lines (observable with
Hinode/EIS) from \citet{2009ApJ...706....1L} with 48 of them shown in the bottom
panel of Figure \ref{fig:tr}. These are lines emitted at cooler
temperatures than observed by SDO/AIA and so we will consider Gaussian
models with centroid temperatures between log$T=5.5$ to $6.5$.

\begin{figure*}\centering
\includegraphics[width=140mm]{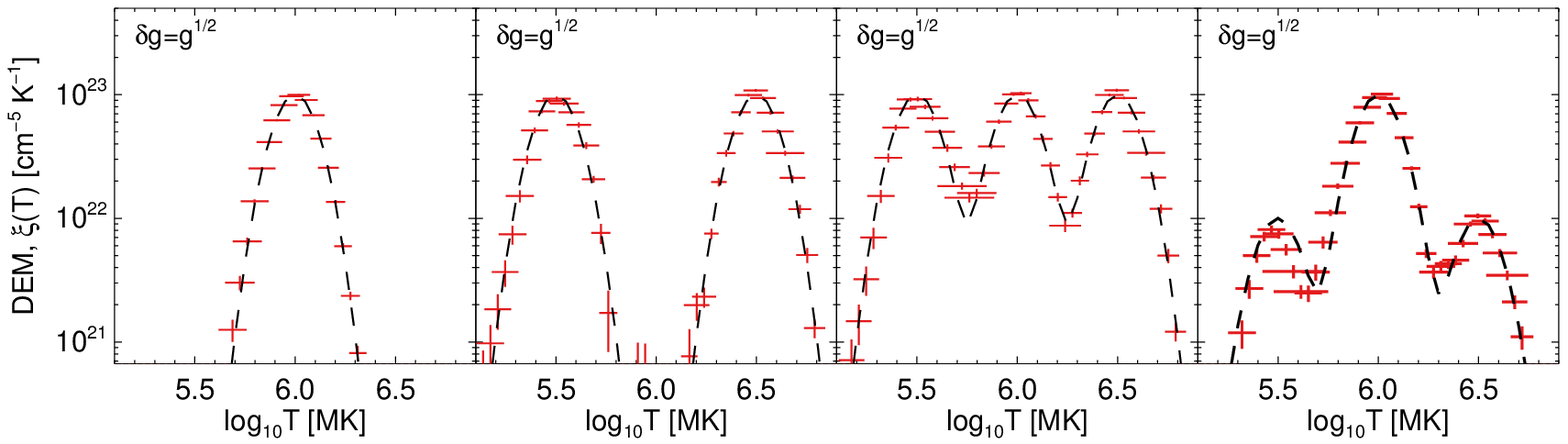}\\
\includegraphics[width=140mm]{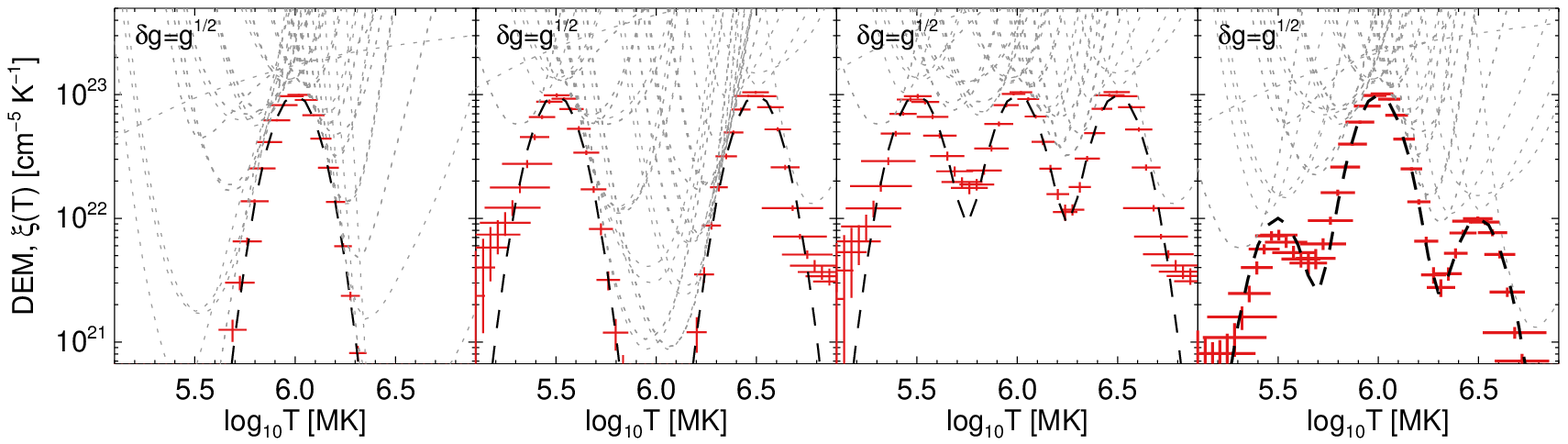}
\caption{\label{fig:eis_mg_poi}DEMs recovered using regularized
inversion (red error bars) from simulated Hinode/EIS data of multiple component
Gaussian models (black dashed lines) of differing $\sigma$ (increasing left to
right). Here the error in the line intensity is taken as the Poisson noise,
which
range between 0.2 to 11\% of the line intensity. In the bottom row the minimum
of the EM loci curves (grey dotted lines) has been used as the guess solution
$\xi_0(T)$.}
\end{figure*}

\begin{figure*}\centering
\includegraphics[width=140mm]{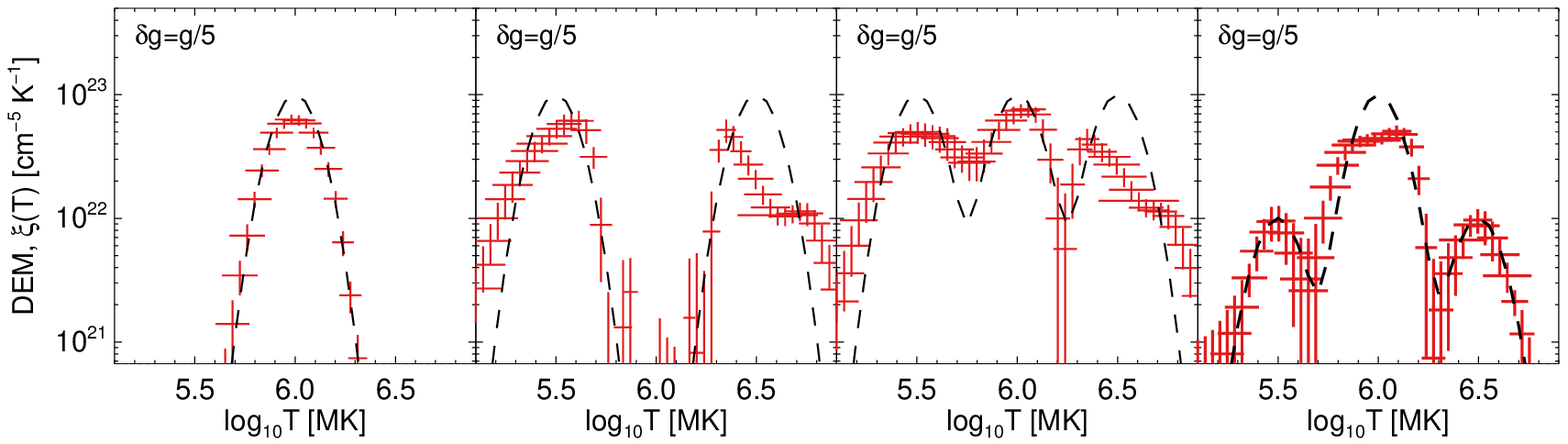}\\
\includegraphics[width=140mm]{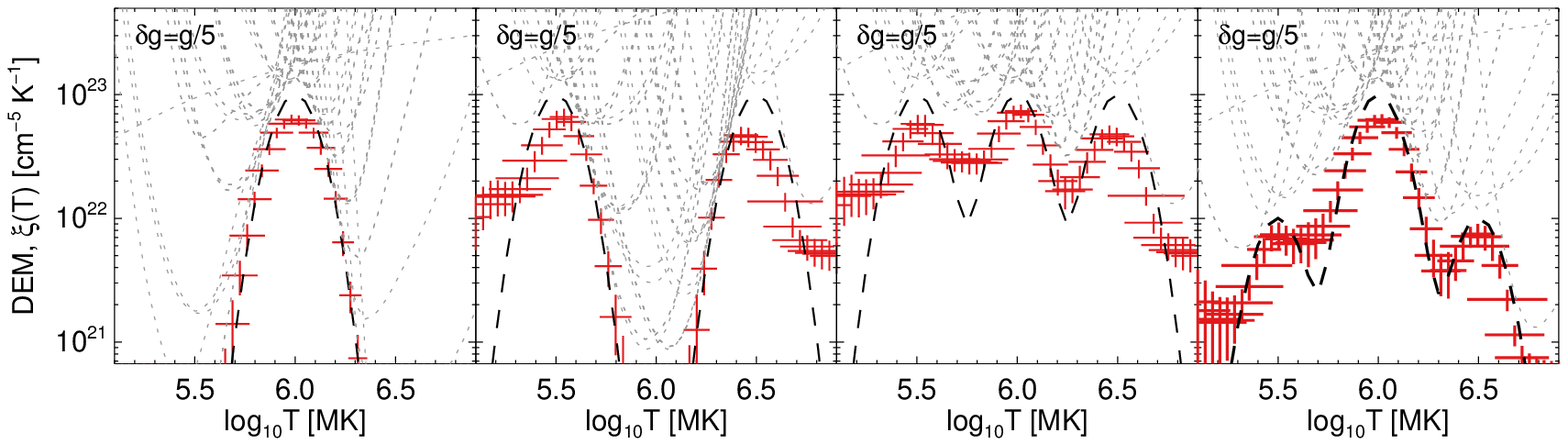}
\caption{\label{fig:eis_mg_per}DEMs recovered using regularized
inversion (red error bars) from simulated Hinode/EIS data of multiple component
Gaussian models (black dashed lines) of differing $\sigma$ (increasing left to
right). Here the error is taken as 20\% of the line intensity. In the bottom row
the
minimum of the EM loci curves (grey dotted lines) has been used as the guess
solution $\xi_0(T)$.}
\end{figure*}

The simulated line intensities are created as before with the pixel
intensities for SDO/AIA but this time we consider not only the Poisson noise
($\delta g_i=\sqrt{g}$) but also a systematic uncertainty that is a percentage
of
each line intensity. This corresponds to a temperature independent factor per
line $c_i$, $\delta g_i=g_i/c_i$. We are using this approach as there are
uncertainties in the relative (few percent), absolute (up to 22 \%) and
modelling
($\approx 10\%$) of the contribution functions for each line (\textit{private
communication P. Young} and \citet{2009ApJ...706....1L}). Such a temperature
independent systematic on the contribution function ${\bf \delta K_i}={\bf
K_i}/c_i$, where $c_i$ is the factor per line, is equivalent to the same
systematic on the observable $\delta g_i=g_i/c_i$, as discussed in
\S\ref{sec:terr}. So when calculating the error in the simulated observable for
our chosen DEM we choose either the Poisson noise or a percentage error, in
the latter case to represent this possibly dominant systematic uncertainty in
the
contribution functions.

\begin{figure*}\centering
\includegraphics[width=140mm]{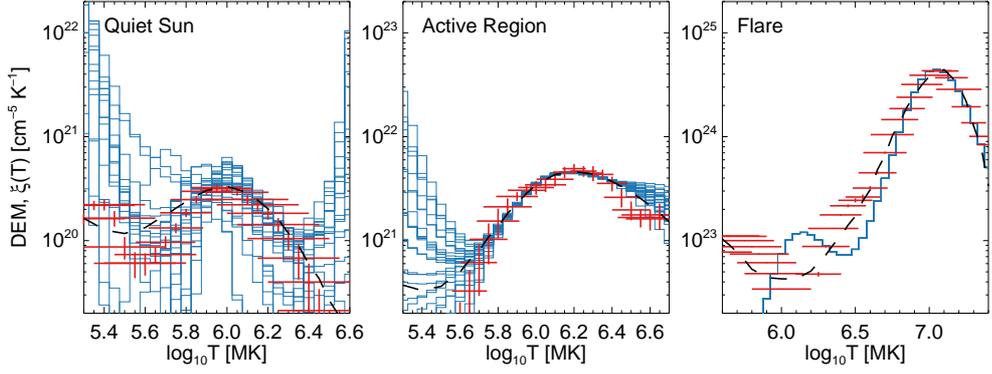}
\caption{\label{fig:aia_ch}Regularized DEMs (red error bars) and
 \texttt{xrt\_dem\_iterative2.pro} solution (blue histograms, with
25 MC realisations) for simulated SDO/AIA data of CHIANTI DEM models
 (dashed black lines, left to right: quiet sun, active region, M-Class
flare).}
\end{figure*}

\begin{figure*}\centering
\includegraphics[width=140mm]{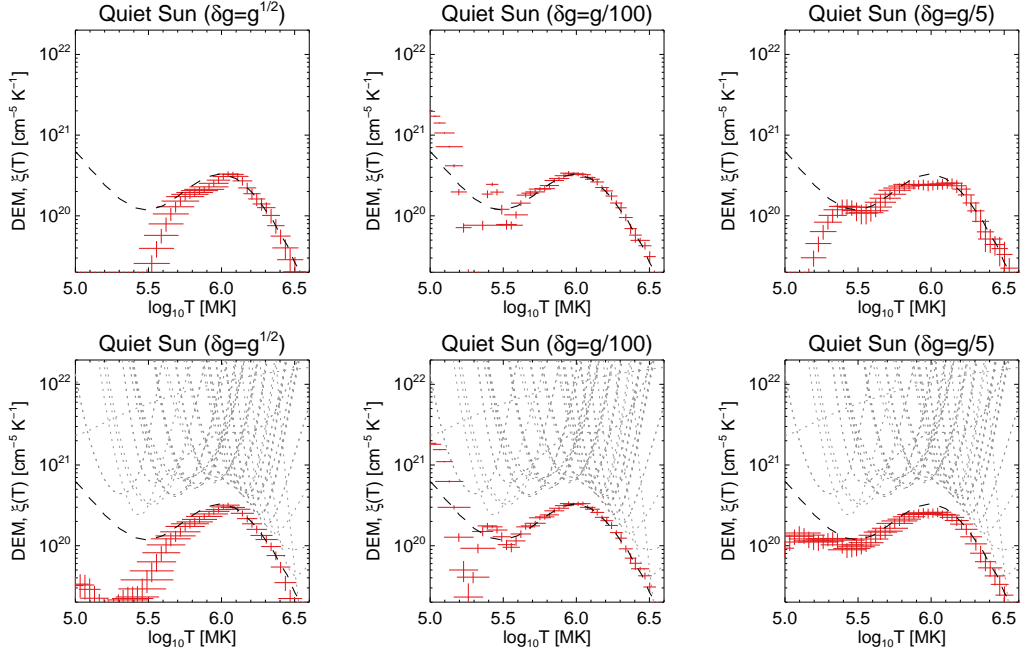}
\caption{\label{fig:eis_qs}Regularized DEMs (red error bars)  for
simulated Hinode/EIS lines of the CHIANTI quiet Sun DEM model. The error is
taken as the Poisson noise and  1\% and 20\% of the line intensity (left to
right).
In the bottom row the minimum of the EM loci curves (grey dotted lines) was used
as the guess solution $\xi_0(T)$.}
\end{figure*}

In Figure \ref{fig:eis_poi} we show the regularized inversion of Gaussian
model DEMs of different widths using a Poisson uncertainty on the line
intensity, again using the zeroth-order constraint and $\alpha=1$ (i.e.
$\chi^2\approx 1$). For the model DEMs used the
uncertainties given by $\delta g_i=\sqrt{g}$ range from 0.2 and 11\% of the line
intensities. The regularized solution very closely matches the model DEM for the
narrowest cases (top left panels), matching the model down to the temperature
resolution of the atomic data used to calculated the contribution functions,
i.e.
$\sigma_T=0.1$. For the wider Gaussian models (top right panels) the
regularized solution recovers the model well about the peak emission but
underestimates the emission at the lowest temperatures. These regularized
DEMs were found using the two-stage approach of no initial guess solution
$\xi_0(T)$ but we can also find the solutions using the minimum of the EM loci
curves (grey dashed lines) as the guess solution (see \S\ref{sec:guess}), shown
in the bottom row of Figure \ref{fig:eis_poi}. For the narrowest Gaussian
models there is little difference in the regularized solutions. The only
improvement here is in the time it takes to perform the computation since the
regularization has only been calculated once. However even with the two-stage
approach the DEM is computed in just a few seconds. For the wider Gaussian
models the regularized DEM found using the guess solution recovers the model
DEM better, particularly at lower temperatures.

\begin{figure*}\centering
\includegraphics[width=140mm]{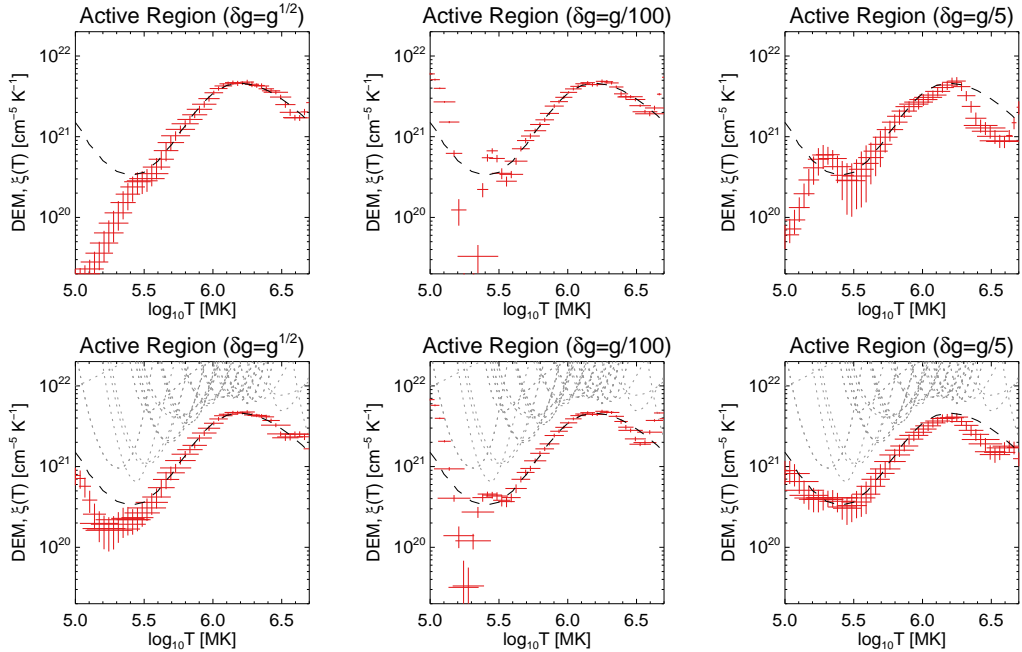}
\caption{\label{fig:eis_ar}Regularized DEMs (red error bars)  for simulated
Hinode/EIS lines of the CHIANTI active region DEM model. The error is taken as
the Poisson noise and  1\% and 20\% of the line intensity (left to right). In
the
bottom row the minimum of the EM loci curves (grey dotted lines) has been
used as the guess solution $\xi_0(T)$.}\end{figure*}

This analysis is repeated but this time the error on the line intensities is
taken to be 20\% instead of the Poisson noise. With this increased uncertainty
the regularization method does not recover the model DEMs as well, shown in
Figure \ref{fig:eis_per}. For the narrowest DEM (top left panel) the regularized
inversion recovers the majority of the model DEM but underestimates the peak
emission and increases its width, the latter showing a reduction in temperature
resolution due to noise. The use of the minimum of the EM loci curves as the
guess solution $\xi_0(T)$ greatly improves the recovery of the source DEMs
(bottom row) but still underestimates the peak emission in the narrowest cases.
This demonstrates that the the temperature resolution had been inherently
degraded by the increase in noise.

We now consider model DEMs constructed of several Gaussian
components of width at the temperature resolution of the atomic data,
$\sigma_T\approx 0.1$, shown with Poisson noise and then 20\% uncertainty
in the line intensity in Figures \ref{fig:eis_mg_poi} and \ref{fig:eis_mg_per}.
With the lower level of noise from the Poisson errors the regularized inversion
recovers well the model DEMs (top row Figure \ref{fig:eis_mg_poi}), even in the
cases with three Gaussian components. The use of the initial guess solution
(bottom row Figure \ref{fig:eis_mg_poi}) generally does worse in recovering the
DEMs at the extremes of the temperature ranges but produces similar results at
the mid-range temperatures where the DEMs peak. With a larger error in the
line intensity the regularized inversion struggles to recover the model DEMs
(Figure \ref{fig:eis_mg_per}). In all cases it is able to recover the correct
number
of distinct components in the DEMs but is unable to match the peak emission,
often producing flatter solutions. The use of the initial guess solution does
help
recover the model DEMs better (bottom row Figure \ref{fig:eis_mg_per}) but it
still struggles with this noisy simulated data.

\section{Simulated Data: CHIANTI Model DEMs}\label{sec:mod_chi}

To test the regularization with more physically realistic DEMs we use
those provided with the CHIANTI atomic database
\citep{1997A&AS..125..149D,2009A&A...498..915D}. We use the DEMs for the
quiet Sun \citep{1999PhDT.........8D}, an active region
\citep{2003A&A...400..737A} and a M2-GOES Class flare
\citep{1979ApJ...229..772D}. With SDO/AIA we test all three of these models
(see \S \ref{sec:aia_chi}) but for the Hinode/EIS lines we only consider the
quiet
Sun and active region models (see \S \ref{sec:eis_chi}) since these lines are
sensitive to temperatures predominantly below log$T\approx 6.5$, lower than
expected in a large flare. Again the regularized solutions were found using the
zeroth-order constraint and as $\alpha=1$ the resulting solutions have
approximately $\chi^2=1$.

\subsection{SDO/AIA Simulated Data}\label{sec:aia_chi}

The regularized DEMs recovered from simulated SDO/AIA data of the
CHIANTI model DEMs is shown in Figure \ref{fig:aia_ch}. The first DEM shown is
for the quiet Sun and the regularization recovers the DEM well but has very large
error bars. This is understandable given the faint, and hence noisy, emission:
the peak emission for the quiet Sun DEM is about $10^{20}$ cm$^{-5}$ K$^{-1}$
which is over three orders of magnitude smaller that the Gaussian examples in
\S \ref{sec:aia_gauss}. In comparison,  \texttt{xrt\_dem\_iterative2.pro} poorly
recovers the model DEM producing a large spread of the MC realisations. The
active region model DEM (middle panel Figure \ref{fig:aia_ch}) is better
recovered than the quiet Sun model which is expected given that this DEM
produces a stronger signal. Both methods have trouble in recovering the model
DEM at the lowest temperatures (log$T<5.7$) where the emission is smallest:
the regularization solution underestimating the emission, the iterative approach
overestimating. For the flare DEM (right panel Figure \ref{fig:aia_ch}) both
methods recover the model well about the peak temperature (log$T\approx
7.0$) will small vertical uncertainties. This is due to the large emission from the
flare, about 3 to 4 orders of magnitude larger than the quiet Sun and active
region models. At lower temperatures the regularized solution matches the
model DEM well though with a substantial horizontal spread. The
\texttt{xrt\_dem\_iterative2.pro} solutions deviate from the model below
log$T\approx6.7$ and produces a false peak about log$T\approx 6.1$ when in
the model it is a minimum.

\begin{figure*}\centering
\includegraphics[width=130mm]{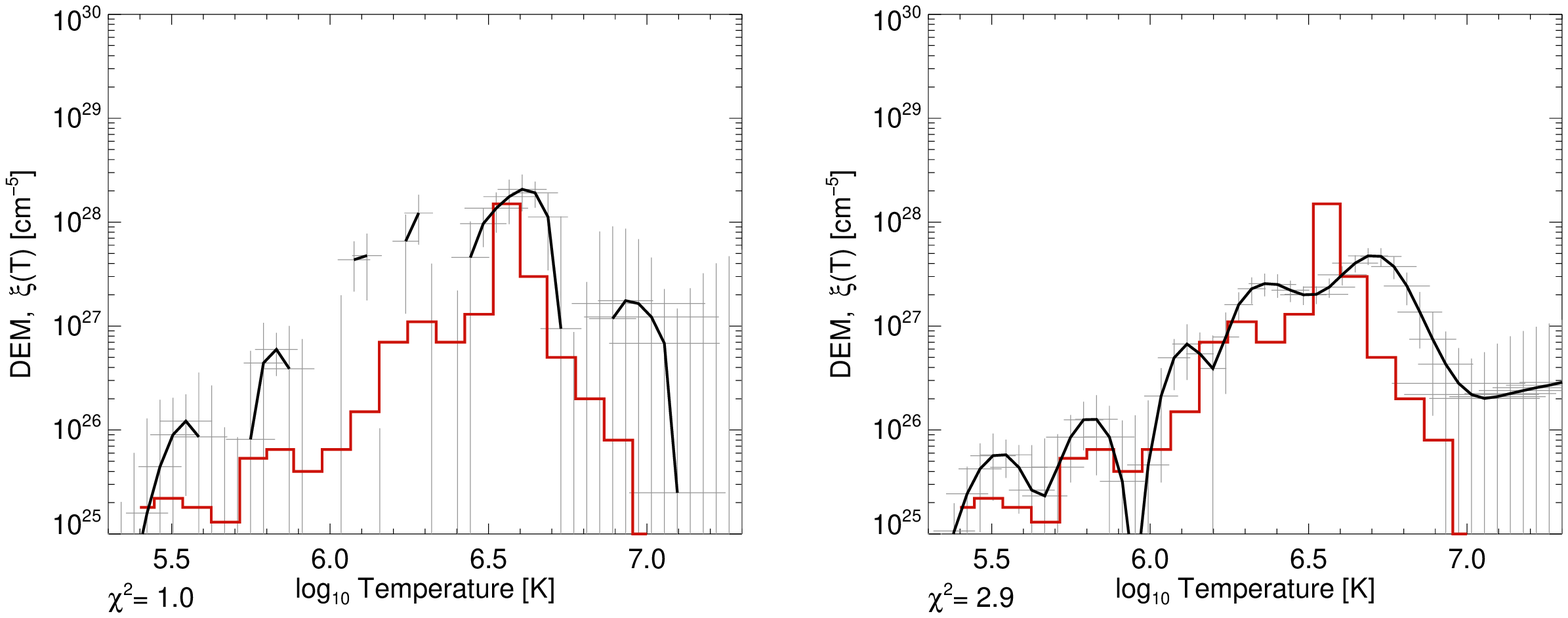}\\
\includegraphics[width=130mm]{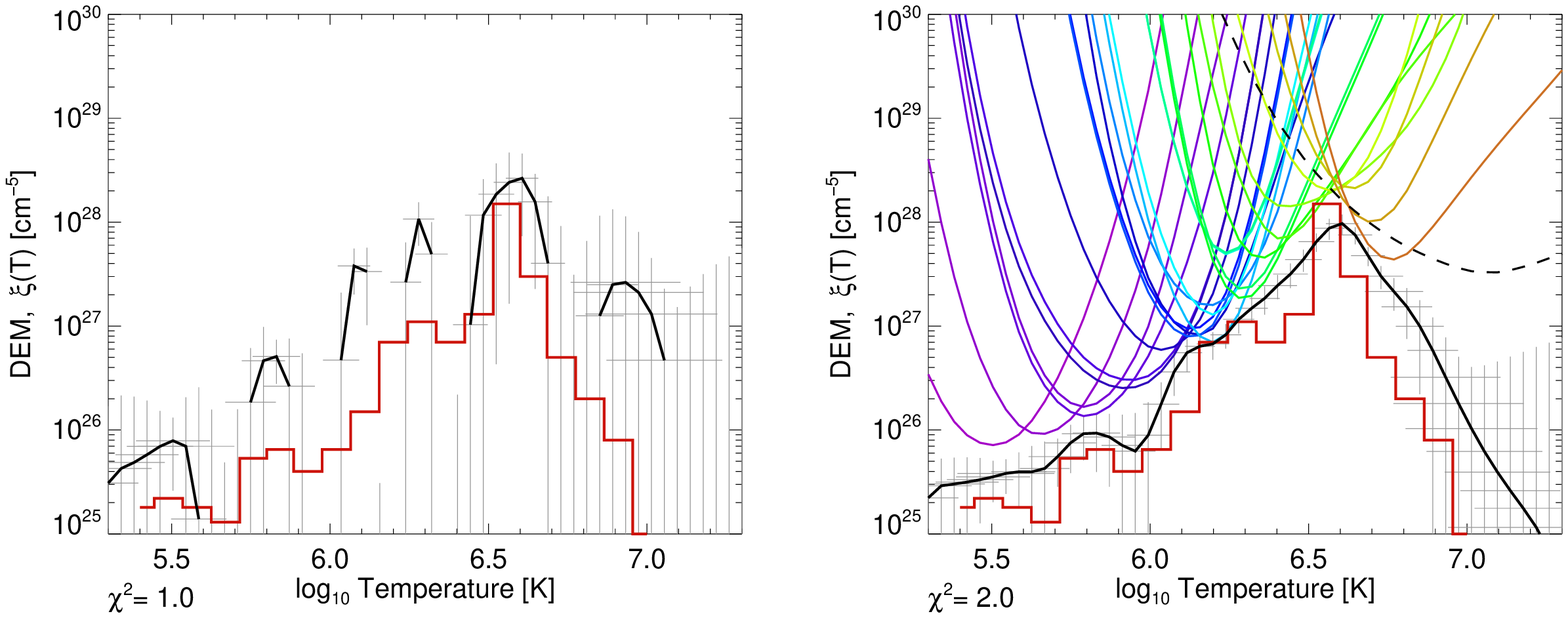}
\caption{\label{fig:real_ar}Regularized emission measure
distributions (black solid line and grey error bars) for the active region core
observations made by \citet{2010ApJ...711..228W}, their MCMC solution (red
histogram) is shown for comparison. In the right-hand panels the positivity
constraint was used and hence larger $\chi^2$ values are achieved by the
regularized solutions. The EM loci curves for the Hinode/EIS (coloured lines) and
Hinode/XRT filter (black dashed line) are shown in the bottom left panel and
their minimum was used as an initial guess solution for the regularizations
shown in the bottom row.}
\end{figure*}

\subsection{Hinode/EIS Simulated Data}\label{sec:eis_chi}

The regularized DEM of the CHIANTI quiet Sun model using the
Hinode/EIS lines is shown in Figure \ref{fig:eis_qs}, where it has been calculated
using a variety of errors in the line intensities and with/without the minimum of
the EM loci as the initial guess solution. When no initial guess solution is used
(top row) the regularization method is able to recover the main peak of emission
(about log$T\approx6.0$) for the three different error cases, Poisson noise, 1\%
and 20\%. Only with the 1\% errors on the line intensities does the regularized
solution properly recover that the emission increases with decreasing
temperature but does not match it very well (middle panel). Although it should
be noted that this model produces very weak emission and the DEM recoverable
from
these multiple Hinode/EIS lines is considerably better than with the SDO/AIA
simulated data (left panel Figure \ref{fig:aia_ch}). When the minimum of the EM
loci
curves is used as the initial guess solution (bottom row, Figure \ref{fig:eis_qs})
the regularized solution does recover more of the model DEM but the problems
with the low temperature component persist.

In Figure \ref{fig:eis_ar} we show the same analysis but this time for the
CHIANTI active region DEM model. The 1\% error case again well recovers the
majority of model DEM but struggles at low and high temperatures when higher,
and more realistic, errors in the line intensities are used. When the minimum of
the EM loci curves are used as the initial guess solution (bottom row, Figure
\ref{fig:eis_ar}) the regularization method does considerably better by
impressively recovering the model DEM even in the situations with larger
errors.

\section{Real Data: Hinode/XRT and EIS}\label{sec:real}

To test the performance of the regularization on real data we use the
observations of an active region core with Hinode/EIS and XRT from
\citet{2010ApJ...711..228W}. This article provides the line intensities, with errors
(which were 20\% to 23\%), for 24 EIS lines and one XRT filter (Al-thick) (see
Table 1, \citet{2010ApJ...711..228W}) and computed the DEM (Figure 4, red
histogram, \citet{2010ApJ...711..228W}) using the MCMC method packaged with
PINTofALE \citep{1998ApJ...503..450K}. Using the information given we use
CHIANTI to calculate the contribution functions for the 24 EIS lines, and with the
Hinode/XRT temperature response function for the Al-thick filter
\citep{2007SoPh..243...63G}, and the quoted intensities and errors we calculate
the zeroth-order regularized solutions and $\alpha=1$ (shown in Figure
\ref{fig:real_ar}). Here we have the regularized solutions found both with and
without the positivity constraint (left vs right columns) and the initial guess
solution from the minimum of the EM loci curves (top vs bottom rows). Shown
for comparison is the MCMC solution found by \citet{2010ApJ...711..228W}. Note
that for all the regularized solutions the resulting ${\bf R_\lambda K}$ matrix
was almost diagonal over log$T\approx 5.5 - 6.9$ indicating that the
regularization has successfully worked over this temperature range. With no
positivity constraint (left-hand panels) the regularized solution is highly
oscillatory between positive and negative values, an indication that this is an
over-regularized solution \citep{1977A&A....61..575C,1985InvPr...1..301B}.
Although, the regularization does produce a maximum at the same temperature
as the MCMC method. The use of the initial guess solution (bottom left panel
Figure \ref{fig:real_ar}) results in minimal changes to the recovered DEM. The
positivity constraint produces a closer match to the MCMC solution. When no
initial guess solution is used (top right panel) a highly oscillatory DEM is
recovered and deviates from the MCMC solution at the peak temperature and
above. When the minimum of the EM loci curves are used as the initial guess
solution (bottom right panel) a smaller $\chi^2$ is achieved while still having a
positive solution. This regularized DEM is a close match to MCMC solution, even
more so when one includes the spread of the DEM found from 250 MC
solutions (shown in Figure 4, \citet{2010ApJ...711..228W}). The crucial
difference though is in the computation time: the regularized solution is found
within a few seconds, the MCMC method taking orders of magnitude longer.

\section{Discussion  \& Conclusions}

In this work, we have applied a regularized inversion technique
developed for RHESSI HXR analysis to multi-filter observations of hot solar
plasma with Hinode/EIS, XRT and SDO/AIA. This method successfully recovers a
variety of model DEMs from different simulated broadband and spectroscopic data
of varying noise and uncertainty. It is implemented\footnote{The code written in
IDL requiring SSW is available online:
http://www.astro.gla.ac.uk/$\sim$iain/demreg/} using General Singular Value
Decomposition (GSVD) and this has several advantages over previous approaches
used to find the regularized inversion of solar data
\citep[i.e][]{1977A&A....61..575C,1997ApJ...475..275J}:
\begin{enumerate}
\item It reliably recovers the DEM even from noisy data and using only the
weakest (yet mostly physically justifiable) constraint, zeroth-order $L_0$;
 \item This method naturally determines the confidence interval for the
regularized solution by calculating both the vertical and horizontal error bars
allowing an objective assessment of the quality of the data, temperature
response functions and DEM. The use of the ${\bf R_\lambda K}$ matrix also
provides
additional information about the robustness of the regularized solution;
\item It is computationally very quick, with DEMs recovered from
SDO/AIA data in about a second or less and a few seconds when a large number of
spectral lines are used, such as with Hinode/EIS. This is crucial considering
the colossal amount of ever increasing data that is being accumulated by SDO,
Hinode and expected from future missions.
\item The recovery of a positive solution can be  guaranteed though this
approach can hide issues with the data and response functions.
\end{enumerate}

\noindent Misconceptions about the problems with inversion
techniques (such as the smoothness criteria) still exist
\citep[e.g.][]{2011arXiv1112.2857L} but the examples shown in this
paper clearly demonstrate that these views are misplaced and our
regularized inversion approach can robustly recover a variety of DEMs. Moreover
the ability to easily determine the temperature uncertainty is a major
improvement over other methods, crucial when trying to test the possibly
isothermal nature of solar plasma.

DEMs are a very useful tool for characterising the temperature distribution of
the corona and trying to reveal the properties of the mechanisms that are
heating the plasma. However there are several caveats to the reconstructed
DEMs which mean they should not be over-interpreted. Any method that
recovers a DEM from a set of real data cannot guarantee that it is the actual
solution given the possible uncertainties/errors in the data and temperature
response functions (either instrumental or in the atomic physics). The use of
more than one method is highly recommended as this would highlight any
issues and artifacts of a particular approach and demonstrate the role of the
uncertainty in the data on the solution. The computational speed of our
regularization method makes it a minor burden to include when using other
techniques and also allows fast exploration of the effect of the uncertainties on
the DEM, as demonstrated for the simulated data in this paper. Even if several
techniques produce the same DEM, with similar uncertainties, this still leaves
the possibility of there being errors in the temperature response functions due
to mis-calibration of the instrument or information in the atomic data being
absent or erroneous. There is a continuous effort to improve the atomic physics
data, through the sterling work of the CHIANTI team and others. At the moment
there are known issues with the SDO/AIA response functions
\citep[e.g.][]{2011ApJ...732...81A} and the effect of this can be tested by
simulating SDO/AIA data using the ``correct''  response but recover the DEM
from this data using the erroneous responses. However if the emission is not
optically thin or not in thermal equilibrium the DEM and response functions are
not appropriate to describe the temperature distribution of the plasma.

All these problems are compounded by the vast quantities of solar data now
available resulting in the current need for a variety of tools to investigate DEMs
quickly, transparently and easily. The use of simulated data is a vital approach
to investigate the role of each of these issues and combined with a
computationally fast algorithm that provides error estimates, such as our
regularization method,  should provide a quicker determination of the reliability
of DEMs and what can be interpreted from them.

\begin{acknowledgements}

This work is supported by a STFC grant (IGH, EPK). Financial support by
the European Commission through the FP7 HESPE Network is gratefully
acknowledged. We would like to thank the referee for their constructive
criticisms that helped us greatly improve this paper.

\end{acknowledgements}


\end{document}